\documentclass[journal, twoside]{IEEEtran}
\usepackage{array}
\usepackage{threeparttable}
\usepackage{float}
\usepackage{graphicx}
\usepackage{color}
\usepackage{makecell}
\usepackage{placeins}
\usepackage{dblfloatfix}
\usepackage{tabularx,colortbl}
\usepackage{amsmath}
\usepackage{amssymb}
\usepackage{balance}
\usepackage{subfigure}
\usepackage{cite}
\usepackage[hyphens]{url}
\usepackage{hyperref}
\hypersetup{colorlinks=true, linkcolor=blue, anchorcolor=blue, citecolor=blue}
\usepackage{booktabs}
\usepackage{multirow}
\usepackage{bm}

\usepackage{enumitem}
\makeatletter

\newcommand{\Rmnum}[1]{\expandafter\@slowromancap\romannumeral #1@}
\makeatother
\renewcommand{\arraystretch}{1.3}

\begin{document}

\bstctlcite{BSTcontrol}

\title{Transfer Learning-Enabled Distortion Compensation for Amplitude-Phase-Time Block Modulation-Based Nonlinear Single-Carrier Wireless Communications} 


\author{
Guoxing~Duan,~\IEEEmembership{Graduate Student Member,~IEEE},
Min~Fan,~\IEEEmembership{Graduate Student Member,~IEEE},\\
Cheng~Yi,~\IEEEmembership{Member, IEEE},
Bensheng~Yang,
Wei Xu, \IEEEmembership{Fellow, IEEE}, \\
Haiming~Wang,~\IEEEmembership{Member,~IEEE},
and~Xiaohu You, \IEEEmembership{Fellow, IEEE}

\thanks{Manuscript received **; revised **; accepted **. Date of publication \qquad\qquad; date of current \qquad. This work was supported in part by the National Natural Science Foundation of China under Grants 62550015 and 62271133, and in part by the Science Foundation of Jiangsu Province of China under Grant BG2025001. \emph{(Corresponding author: Haiming Wang and Wei Xu)}}
\thanks{Guoxing Duan, Min Fan, Cheng Yi, and Haiming Wang are with the School of Information Science and Engineering and the State Key Laboratory of Millimeter Waves, Southeast University, Nanjing 211189, China, and also with the Pervasive Communication Research Center, Purple Mountain Laboratories, Nanjing 211111, China (e-mail: duanguoxing@seu.edu.cn; minfan@seu.edu.cn; yicheng@pmlabs.com.cn; hmwang@seu.edu.cn).} 
\thanks{Bensheng Yang is with the College of Telecommunications and Information Engineering, Nanjing University of Posts and Telecommunications, Nanjing 210003, China (email: yangbensheng@njupt.edu.cn).}
\thanks{Wei Xu and Xiaohu You are with the School of Information Science and Engineering and the National Mobile Communications Research Laboratory, Southeast University, Nanjing 211189, China, and also with the Pervasive Communication Research Center, Purple Mountain Laboratories, Nanjing 211111, China (e-mail: wxu@seu.edu.cn; xhyu@seu.edu.cn).}
\thanks{Color versions of one or more of the figures in this paper are available online at http://ieeexplore.ieee.org.}
\thanks{Digital Object Identifier \qquad}
}

\markboth{IEEE Transactions on Communications}{Duan \MakeLowercase{\textit{et al.}}: TL-Enabled Distortion Compensation for APTBM-Based Nonlinear Single-Carrier Wireless Communications}

\maketitle

\boldmath
\begin{abstract}
Power amplifier (PA) nonlinearity and memory effects significantly limit the spectral compliance, reliability, and energy efficiency of communication systems. To address this, we propose a transfer-learning-enabled, fully digital transceiver-cooperative method for amplitude-phase-time block modulation (APTBM)-based nonlinear single-carrier transmission under adjacent channel leakage ratio (ACLR) constraints. At the transmitter, iterative clipping and filtering (ICAF) and static digital pre-distortion (SDPD) act jointly to reduce signal peaks and suppress spectral regrowth without requiring wideband feedback. At the receiver, the inherent amplitude–phase constraints of APTBM provide weakly supervised prior knowledge for offline inverse-model pretraining, which is followed by the online few-shot adaptation of a lightweight digital post-distortion (DPoD) network. Subsequently, a cascaded DPoD and clipping-noise cancellation scheme systematically compensates for residual distortions induced by both the PA and ICAF. Simulation and measurement results demonstrate reliable transmission at an input back-off of approximately 2 dB under a 30-dBc ACLR constraint. Furthermore, the proposed DPoD approach significantly reduces online training time and computational overhead, delivering a performance gain of over 2 dB compared to conventional learning-based DPoD schemes.
\end{abstract}
\unboldmath

\section{Introduction}
\label{Sec_Int}

\IEEEPARstart{W}{ith} the large-scale deployment of fifth-generation (5G) base stations and the evolution toward sixth-generation (6G) networks, the energy consumption of wireless communication systems continues to rise, making energy efficiency (EE) an increasingly critical design objective \cite{ZhongGe2024_6GCarbonNeutral,YangEtAl2025_6GSustainableMobile}. As one of the most energy-intensive components in wireless transmitters, power amplifiers (PAs) exhibit an inherent trade-off between power-added efficiency (PAE) and linearity, which becomes particularly pronounced under high-frequency, wideband, and high-order modulation operation \cite{CamarchiaEtAl2020_mmWavePAReview}. 

Operating a PA near saturation improves output power and PAE but introduces amplitude modulation (AM)–AM and AM–phase modulation (PM) distortion, resulting in in-band error vector magnitude (EVM) deterioration and out-of-band (OOB) spectral regrowth \cite{CamarchiaEtAl2015_DohertyReview,ZhangEtAl2020_First20YearsGreenRadios,HanEtAl2011_GreenRadioTechniques}. The former can be mitigated at the receiver, whereas the latter directly violates adjacent channel leakage ratio (ACLR) requirements and interferes with adjacent channels. To mitigate PA-induced performance degradation, existing approaches include input back-off (IBO) \cite{AzoliniTavaresEtAl2016_IBOOptimizationOFDM}, peak-to-average power ratio (PAPR) reduction \cite{RahmatallahMohan2013_PAPROFDM_Survey}, digital pre-distortion (DPD) \cite{HaiderEtAl2022_PredistortionSurvey_mmWave}, and digital post-distortion (DPoD) \cite{OrabiEtAl2025_PostdistortionAlternative}. 

Among these, IBO mitigates nonlinear distortion but directly sacrifices PAE \cite{AzoliniTavaresEtAl2016_IBOOptimizationOFDM}. PAPR-reduction methods alleviate the required back-off by reshaping or selecting low-peak waveforms \cite{WangEtAl2020_JointCFRDPD_ClippingBankFiltering,SunOchiai2021_ClippedFilteredOFDM_IDR,HouEtAl2010_NonlinearCompandingPAPR,DeumalEtAl2011_CubicMetricTR,KouLuAntoniou2007_ConstellationExtensionPAPR,ShiEtAl2024_SLM_MixedNumerologyNOMA,KuWangChen2010_ReducedComplexityPTS}. Among them, iterative clipping and filtering (ICAF) is attractive for its low implementation complexity, although its clipping operation introduces irrecoverable in-band distortion if no receiver-side compensation is applied \cite{WangEtAl2020_JointCFRDPD_ClippingBankFiltering,SunOchiai2021_ClippedFilteredOFDM_IDR}. Other techniques commonly incur spectral efficiency (SE) loss, require side information, or increase baseband complexity. More importantly, PAPR reduction alone cannot compensate for the PA distortion or guaranty ACLR compliance near saturation \cite{WangEtAl2020_JointCFRDPD_ClippingBankFiltering}.

In contrast, DPD is a widely adopted transmitter-side solution that applies an inverse behavioral model to predistort the PA input signal, thereby simultaneously improving in-band linearity and suppressing spectral regrowth \cite{HaiderEtAl2022_PredistortionSurvey_mmWave}. Classical memory-polynomial (MP) and generalized memory-polynomial (GMP) models provide effective descriptions of nonlinear and memory-dependent PA behavior \cite{MorganEtAl2006_GMP}. More recently, neural network-based DPD methods, such as real-valued focused time-delay neural networks (RVFTDNN) and augmented real-valued time-delay neural networks (ARVTDNN), have further improved modeling capability under wideband and strongly nonlinear operation \cite{Rawat2012_RVFTDNN,WangEtAl2019_ARVTDNN}. Additionally, in high-dynamic and integrated sensing-and-communication scenarios, DPD schemes based on neural networks also demonstrate superior performance gains \cite{YuEtAl2025_DPD_mmWaveGaN_6G_ISAC_SWIPT,ZhaoEtAl2025_ARVTDform}. However, conventional DPD relies on a high-speed feedback observation path with wideband sampling capability and accurate path calibration, while requiring online model adaptation to track PA variations \cite{LiWangZhu2020_SamplingRateReductionDPD,Wood2017_SystemLevelDPD}. Such feedback-assisted architectures introduce additional hardware costs, calibration complexity, and power consumption, limiting their applicability to compact terminals, integrated transceivers, satellite links, and rapidly varying operating conditions \cite{Wood2017_SystemLevelDPD}.

DPoD provides an alternative strategy by moving part of the nonlinear compensation burden from the transmitter to the receiver \cite{OrabiEtAl2025_PostdistortionAlternative}. Leveraging the greater computational resources and flexible signal processing capabilities of receivers, DPoD enables effective distortion compensation without requiring feedback hardware or complex real-time processing at the transmitter. Early receiver-side nonlinear mitigation schemes include PA nonlinear cancellation (PANC) \cite{TelladoHooCioffi2003_MLDetectionNonlinearOFDM} and the reconstruction of distorted signals (RODS) \cite{AlinaAmrani2016_OnDigitalPostDistortion}. More recent studies have investigated a sum-of-products PA representation-based least-squares equalizer \cite{HeEtAl2024_UnifiedPARepresentationReceiverEQ} and deep-learning-based receivers \cite{SalmanGuvensen2021_QAMDetector_NonlinearDPoD_FDEBank, PihlajasaloEtAl2023_DeepLearningOFDMReceiversPowerEfficiency} for improved coverage and EE. Additionally, in \cite{LiuEtAl2025_MetaLearningReceiver_MIMO_OFDM_HPA}, a meta-learning-based adaptive compensation scheme was also proposed for combined offline and online training. Nevertheless, DPoD schemes are strictly confined to scenarios with relaxed ACLR constraints and fail to suppress transmitter-side OOB emission control. In addition, although learning-based methods eliminate the need for a priori PA models, their online training burden remains a critical bottleneck; requiring prolonged slow-time (ST) training sequences inherently degrades SE, while intensive online iterations considerably increase communication latency.

Different from previous PA nonlinearity mitigation strategies, amplitude-phase-time block modulation (APTBM) is proposed in \cite{Fan2025_APTBM} for nonlinear distortion signal reconstruction. This modulation maps information into symbol-blocks formed by two adjacent modulated symbols in the time domain, imposing amplitude and phase constraints. The block-level constraints enable model-agnostic receiver reconstruction, but existing reconstruction remains inaccurate under strong nonlinearity, memory effects, and high-order modulation. The reconstruction performance depends strongly on the AM–AM distortion characteristics, and errors may be amplified by inaccurate phase-term estimation, leading to a stepwise deterioration in the distortion compensation performance. The optimization-based extension in \cite{Xia2026_TwoStageAPTBM} further requires accurate PA phase-distortion knowledge and still does not address ACLR compliance.

The aforementioned transmitter-side compensation methods ensure PA linearity at the cost of high hardware complexity and online adaptation overhead, while receiver-side DPoD reduces the transmitter burden but cannot suppress OOB emission or provide reliable prior knowledge for online compensation. To address these issues, this paper proposes a transfer learning-enabled fully digital transceiver-cooperative framework for nonlinear single-carrier transmission. The transmitter employs ICAF and static DPD (SDPD) \cite{Fan2026_APFBM} to satisfy ACLR requirements and lower IBO. The receiver exploits the amplitude–phase structural prior of APTBM to enable offline weakly supervised pretraining and online few-shot adaptation. This separation of OOB leakage control and in-band distortion compensation allows the PA to operate more efficiently under strict spectral constraints, achieving higher PAE with reduced resource overhead while maintaining reliable information recovery. The contributions of this paper are summarized below.

\begin{enumerate} 
    \item A fully digital transceiver-cooperative framework is developed for ACLR-constrained nonlinear single-carrier transmission. Transmitter-side ICAF and SDPD not only reduce PAPR but also suppress OOB emissions by a wide margin, while receiver-side compensation restores in-band linearity without wideband feedback or additional RF hardware.
    \item Transfer learning-enabled DPoD is proposed to reduce transceiver resource overhead by leveraging APTBM constraints as prior knowledge for offline pretraining and adapting a lightweight inverse network using few-shot online samples. It achieves comparable compensation performance with more than 20-fold reductions in training and computational overhead across diverse PA models and high-order modulation schemes.
    \item The cascaded DPoD and clipping-noise  cancellation (CNC) scheme is proposed to sequentially compensate for the coupled distortions induced by ICAF and the PA. Measured and simulation results demonstrate that the staged design mitigates the performance degradation associated with joint compensation, thereby further improving EVM and bit error rate (BER) performance.
\end{enumerate}

The remainder of this paper is organized as follows. Section \ref{Sec_transmissionScheme} introduces the nonlinear wireless transmission model. Sections \ref{Sec_DPoDframework} and \ref{Sec_DPoDframework} present the proposed transfer-learning-enabled DPoD and the cascaded ICAF-PA distortion compensation schemes, respectively. Section \ref{Sec_Exper_Anal} presents and analyzes the results from both simulations and experimental measurements. Finally, Section \ref{Sec_Con} concludes the paper.

\emph{Notation:} Boldface lowercase and uppercase letters denote vectors and matrices, respectively, while calligraphic letters denote sets or symbol alphabets. The symbols $\circledast$ and $\forall$ represent convolution and the universal quantifier, respectively. The operators $I\{\cdot\}$, $Q\{\cdot\}$, $|\cdot|$, $\|\cdot\|_{2}^{2}$, and $\angle(\cdot)$ denote the real component, imaginary component, amplitude, squared Euclidean norm, and phase, respectively. Moreover, $\mathbb{R}$ and $\mathbb{R}^{m\times n}$ denote the real field and the set of $m\times n$ real-valued matrices, respectively.

\section{Nonlinear Wireless Transmission Model}
\label{Sec_transmissionScheme}

In this section, the APTBM symbol-block representation and constraint characteristics are first introduced. Subsequently, the nonlinear transmission model with transceiver cooperation, as well as the ICAF and SDPD processing schemes for the transmitter, is presented.

\subsection{APTBM-Based Symbol Block Representation}
\label{SubSec_APTBM}

Different from conventional modulation schemes that convey information using the absolute amplitude and phase of individual symbols, APTBM maps information into symbol-blocks composed of two consecutive complex symbols in the time domain \cite{Fan2025_APTBM}. Let $N$ denote the number of transmitted modulated symbol-blocks, where $n=1,2,\ldots,N$. The $n$-th symbol-block is represented as 
\begin{equation}
\label{Eq_symbolBlock}
\mathbf{x}_n=\begin{bmatrix}
x_{n,\mathrm{a}},x_{n,\mathrm{b}}
\end{bmatrix}^\mathrm{T},
\end{equation}
where $x_{n,\mathrm{a}}=\left|x_{n,\mathrm{a}}\right|e^{j\psi_{n,\mathrm{a}}}$, $x_{n,\mathrm{b}}=\left|x_{n,\mathrm{b}}\right|e^{j\psi_{n,\mathrm{b}}}$. 

\begin{figure}[tp]%
\centering
\includegraphics[width=3.4in,trim=9mm 8mm 8mm 6mm, clip]{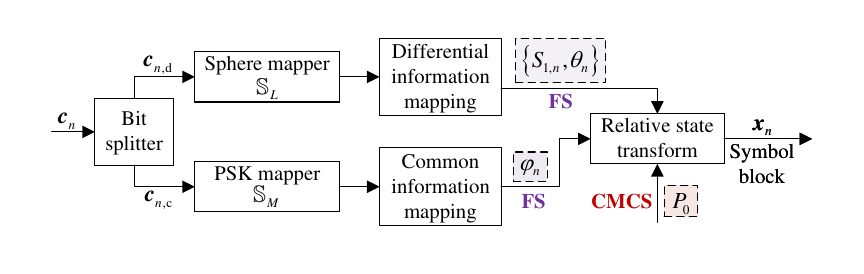}
\caption{Block diagram of the APTBM.}
\label{fig_APTBM_flowchart}
\vspace{-6 pt}
\end{figure}

To characterize the block-wise structure, the four relative state parameters of the symbol-block defined from the Stokes vectors \cite{HenarejosPerezNeira2018_3DPolarizedModulation} are represented as 
\begin{equation}
\label{Eq_four_relativeState}
\begin{cases}
\begin{aligned}
P_n &=\mid x_{n,\mathrm{a}}\mid^{2}+\mid x_{n,\mathrm{b}}\mid^{2}, \\
S_{1,n} &= \mid x_{n,\mathrm{a}}\mid^{2}-\mid x_{n,\mathrm{b}}\mid^{2}, \\
\varphi_{n}&=\psi_{n,\mathrm{a}}+\psi_{n,\mathrm{b}}, \\
\theta_{n}&=\psi_{n,\mathrm{a}}-\psi_{n,\mathrm{b}},
\end{aligned}
\end{cases}
\end{equation}
where $P_n$ represents the power sum of the symbol-block, $S_{1,n}$ and $\theta_{n}$ denote the power difference and phase difference, respectively, and $\varphi_{n}$ is the phase sum. The complex expression of $x_{n,\mathrm{a}}$ and $x_{n,\mathrm{b}}$ in the symbol-block is derived from the transformation of four relative state parameters \cite{Fan2025_APTBM} as 
\begin{equation}
\label{Eq_symbolBlock_complex}
\begin{cases}
x_{n,\mathrm{a}}=e^{-j\frac{\varphi_n}{2}}\sqrt{\frac{P_n+S_{1,n}}{2}}e^{-j\frac{\theta_n}{2}}, \\
x_{n,\mathrm{b}}=e^{-j\frac{\varphi_n}{2}}\sqrt{\frac{P_n-S_{1,n}}{2}}e^{j\frac{\theta_n}{2}}.
\end{cases}
\end{equation}

To mitigate PA nonlinear distortion induced by inter-symbol amplitude fluctuations, APTBM constrains the total power of each symbol-block to a constant value 
\begin{equation}
\label{Eq_symbolBlock_PowerSum}
P_n=P_0, \forall n.
\end{equation}
This serves as the constant-modulus constraint state (CMCS), while the remaining parameters $\left\{S_{1,n},\theta_n,\varphi_n\right\}$ are utilized as free states (FS) for information carriage. As shown in (\ref{Eq_symbolBlock_complex}), $\varphi_n$ forms the common phase term $e^{-j\varphi_n/2}$ in the symbol-block $\mathbf{x}_n$, while $\left\{S_{1,n},\theta_n\right\}$ characterizes the amplitude and phase of the symbol-block in differential form.

Based on the analysis above, the APTBM symbol alphabet $\mathbb{S}_{LM}$ can be decomposed into a common alphabet $\mathbb{S}_{M}$, which carries $\varphi_{n}$, and a differential alphabet $\mathbb{S}_{L}$, which carries $\left\{S_{1,n},\theta_n\right\}$. The modulation order (MO) of APTBM is configured as $M \times L$, where $M$ and $L$ denote the MOs of $\mathbb{S}_{M}$ and $\mathbb{S}_{L}$, respectively, carrying $k_{L}=\log_2(L)$ and $k_{M}=\log_2(M)$ bits. Thus, the number of bits carried by the $n$-th symbol-block is $k=k_{L}+k_{M}$.

Fig. \ref{fig_APTBM_flowchart} illustrates the mapping process of the bit and symbol-blocks in APTBM. After channel coding, the bit vector $\mathbf{c}_n$ is divided into two subvectors, $\mathbf{c}_{n,\mathrm{c}}$ and $\mathbf{c}_{n,\mathrm{d}}$, according to the ratios $k_{M}$ and $k_{L}$. The two subvectors are then mapped by $\mathbb{S}_{M}$ and $\mathbb{S}_{L}$ to carry common and differential information, respectively. By combining the CMCS parameter ($P_0$), the four relative states of the symbol-block are constructed. The complex representation of the symbol-block $\mathbf{x}_n$ in (\ref{Eq_symbolBlock_complex}) is obtained through parameter transformation.
 
Under the constant-modulus constraint of each symbol-block, variations in the common phase do not induce amplitude fluctuations and are therefore insensitive to PA nonlinear distortion. To ensure noise resistance while maintaining inherent robustness to the nonlinearity of PA, the mapping from $\mathbf{c}_{n,\mathrm{c}}$ to $\varphi_n$ is performed using constant modulus phase shift keying (PSK). Thus, $\varphi_n$ is given by $\frac{\varphi_n}{2}\in\mathbb{S}_M$, where
\begin{equation}
\label{Eq_PSK}
\mathbb{S}_M=\left\{\left.\frac{2\pi(m-1)}{M}\right|m=1,2,\ldots,M\right\}.
\end{equation}
Furthermore, to satisfy the constant-modulus constraint and maximize the minimum Euclidean distance (MED) among elements of $\mathbb{S}_L$, the set $\mathbb{S}_L$ is constructed by selecting $L$ state points that are approximately uniformly distributed on the unit sphere using a Fibonacci point set. The unit sphere is parameterized by the power-normalized spherical coordinates $\begin{Bmatrix} \tilde{S}_{1,l},\tilde{S}_{2,l},\tilde{S}_{3,l} \end{Bmatrix}$, where $l=1,2,\ldots,L$. Each element $\left\{S_{1,l},\theta_l\right\}$ in $\mathbb{S}_L$ is represented as $\theta_l=\arctan(\frac{\tilde{S}_{3,l}}{\tilde{S}_{2,l}})$, $S_{1,l}=\tilde{S}_{1,l}$. Thus, for the $n$-th APTBM symbol-block, the differential information satisfies $\left\{S_{1,n},\theta_n\right\}\in\mathcal{S}_L$ \cite{Fan2025_APTBM}.

It is worth noting that the constant power-sum constraint of each APTBM symbol-block not only contributes to a lower PAPR modulation waveform \cite{Fan2025_APTBM}, but also provides useful prior knowledge for the receiver-side reconstruction of PA-induced nonlinear distortion.

\begin{figure*}[tp]%
\centering
\includegraphics[width=6.8in,trim=6mm 6mm 6mm 6mm, clip]{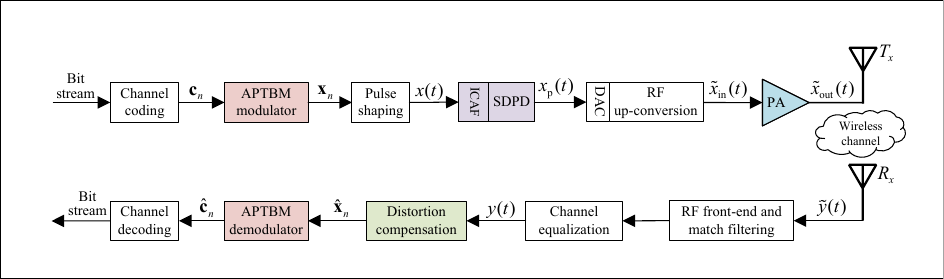}
\caption{APTBM-based nonlinear wireless transmission scheme with ACLR-constrained transmitter processing and receiver-side distortion compensation.}
\label{fig_WirelessCommScheme}
\vspace{-6 pt}
\end{figure*}

\subsection{Nonlinear Transmission Scheme}
\label{SubSec_transmissionModel}

Based on the APTBM symbol-block representation, the nonlinear wireless transmission scheme is illustrated in Fig. \ref{fig_WirelessCommScheme}. Let $\mathcal{F}(\cdot)$ denote the mapping function from the bit sequence $\mathbf{c}_n$ to the symbol-block $\mathbf{x}_n$. The mapping of $\mathbf{x}_n$ via the alphabet $\mathbb{S}_{LM}$ can be represented as $\mathbf{x}_n=\mathcal{F}(\mathbf{c}_n)$. Following pulse shaping, ICAF, SDPD, digital-to-analog conversion (DAC), and RF up-conversion, the PA input signal can be expressed as:
\begin{equation}
\label{Eq_basebandProce}
\tilde{x}_\mathrm{in}(t)=e^{j2\pi f_ct}
\mathcal{G}_{s}\left\{
\mathcal{T}_{c}\left\{
x(t)
\right\}
\right\},
\end{equation}
where
\begin{equation}
\label{Eq_PulseShaping}
x(t)=\sum_{n=1}^{N}\left[x_{n,\mathrm{a}}p(t_n)+x_{n,\mathrm{b}}p(t_n-T)\right],
\end{equation}
$f_{c}$ is the carrier frequency, $t_n=t-2nT$, and $p(t)$ is the baseband pulse-shaping filter with a symbol period $T$. $\mathcal{T}_{c}\{\cdot\}$ and $\mathcal{G}_{s}\{\cdot\}$ denote the preprocessing operations of ICAF and SDPD on the baseband signal, respectively. The PA behavior is modeled as a nonlinear operator $\mathcal{L}\{\cdot\}$, yielding the output signal 
$\tilde{x}_{\mathrm{out}}(t)=\mathcal{L}\{\tilde{x}_{\mathrm{in}}(t)\}$.
After propagation through the wireless channel $h(t)$ and the addition of complex additive white Gaussian noise $z(t)$, the received signal is given by
\begin{equation}
\label{Eq_channelPropagation}
\tilde{y}(t)=\tilde{x}_{\mathrm{out}}(t)\circledast{h(t)}+z(t).
\end{equation}

At the receiver, the baseband signal $y(t)$ is obtained through the RF front-end, which performs down-conversion and analog-to-digital conversion (ADC), followed by matched filtering and channel equalization. Sampling at the symbol period $t = nT$ yields the distorted symbol-blocks 
\begin{equation}
\label{Eq_Rx_symbolBlock}
\mathbf{y}_n=\begin{bmatrix}
y_{n,\mathrm{a}},y_{n,\mathrm{b}}
\end{bmatrix}^\mathrm{T}.
\end{equation}
Except for ICAF-induced distortion and PA nonlinearity, other channel impairments are assumed to be perfectly compensated. Accordingly, the residual distortion in $\mathbf{y}_n$ and $y(t)$ is dominated by the cascaded ICAF and PA nonlinear distortions.

The objective of the receiver distortion compensation module is to reconstruct the distorted signal by performing the inverse process of the nonlinear behavior of the ICAF and PA. Let $\mathcal{L}^{-1}(\cdot)$ and $\mathcal{T}^{-1}_{c}\{\cdot\}$ denote the inverse mapping process of PA and ICAF, respectively. The reconstruction process for the distorted signal $y(t)$ is then described by
\begin{equation}
\label{Eq_Rx_DistortionCompe}
\hat{{x}}(t)=\mathcal{T}^{-1}_{c}\{
\mathcal{L}^{-1}\{{y}(t)\}
\}.
\end{equation}
Finally, the compensated modulation symbol-block $\hat{\mathbf{x}}_n$ is obtained by downsampling the reconstructed signal $\hat{x}(t)$. It is processed by the maximum-likelihood detection and the APTBM inverse demapping function $\mathcal{F}^{-1}(\cdot)$ to recover the bit vector $\hat{\mathbf{c}}_n$ and the transmitted bit stream.

\subsection{ACLR-Constrained Transmitter-Side Baseband Processing}
\label{SubSec_ICAF_SDPD}

To achieve high PAE, the PA must operate near its saturation region, exacerbating AM-AM and AM-PM distortion, causing spectral regrowth and ACLR degradation. To avoid the PAE loss of IBO and the feedback overhead of conventional DPD, SDPD is adopted as a solution that requires only a few PA parameters to satisfy the ACLR constraint \cite{Fan2026_APFBM}. However, SDPD remains limited by the high PAPR of the baseband signal, since large-envelope samples can still push the PA toward saturation. Thus, ICAF is introduced before SDPD to clip signal peaks and filter OOB clipping components, thereby reducing the PA input PAPR. This enables a higher average PA input power under the same ACLR constraint, improving both output power and PAE.

Specifically, the baseband signal $x(t)$ is clipped to a predefined threshold $A$. The clipped signal $\bar{x}(t)$ is given by
\begin{equation}
\label{Eq_Clip}
\bar{x}(t)=
\begin{cases}
A e^{j\angle(x(t))}, & |x(t)|>A,\\
x(t), & |x(t)|\leq A.
\end{cases}
\end{equation}
The clipping ratio (CR) $\gamma$ in dB is defined as $\gamma = 20\log10(\frac{A}{\sqrt{P_{\mathrm{av}}}})$, where $P_{\mathrm{av}}$ is the average power of the signals before clipping. To suppress spectrum leakage, a Kaiser window-based filter is used to further process the clipped signal, with an OOB attenuation factor set to 30 dB.

Letting $x_{\mathrm{f}}(t)$ and $x_{\mathrm{p}}(t)$ denote the input and output signals of SDPD, respectively, the input amplitude and phase are predistorted based on the pre-characterized static AM-AM and AM-PM responses of the PA \cite{Fan2026_APFBM}. The amplitude pre-distortion is first performed by determining the predistorted envelope $|x_{\mathrm{p}}(t)|$ as
\begin{equation}
\label{Eq_SDPD_amp}
\kappa |x_{\mathrm{f}}(t)||x_{\mathrm{p}}(t)|^{\iota_0\lambda} - |x_{\mathrm{p}}(t)| + |x_{\mathrm{f}}(t)| = 0,
\end{equation}
where $\kappa$ and $\lambda$ are the static AM-AM response parameters determined by $P_{\mathrm{in,sat}}$ and $P_{\mathrm{in,1dB}}$, and $\iota_0=20/\ln 10$. $P_{\mathrm{in,sat}}$ and $P_{\mathrm{in,1dB}}$ represent the PA input power at saturation and 1 dB of compression, respectively. This equation is solved to obtain $|x_{\mathrm{p}}(t)|$, such that the PA output amplitude approaches the desired linear response. Then, the input phase is inversely compensated according to the static AM-PM response of the PA as
\begin{equation}
\label{Eq_SDPD_phase}
\angle(x_{\mathrm{p}}(t)) = 
\angle(x_{\mathrm{f}}(t)) - 
\frac{2\delta_{\psi}}{1+\exp\left[-\lambda_p\left(P_{\mathrm{in}}-P_\mathrm{m}\right)\right]},
\end{equation}
where $\delta_{\psi}$ denotes the average nonlinear phase shift, $P_\mathrm{m}$ is the input power corresponding to the phase-transition center, $\lambda_p$ is the phase-transition rate, and $P_{\mathrm{in}}$ is the instantaneous input power associated with $|x_{\mathrm{p}}(t)|$, as detailed in \cite{Fan2026_APFBM}. Accordingly, the ICAF and SDPD cascaded predistorted signal is constructed as $x_{\mathrm{p}}(t)=|x_{\mathrm{p}}(t)|e^{j\angle(x_{\mathrm{p}}(t))}$.

\section{Transfer Learning-Enabled Receiver-Side Distortion Compensation}
\label{Sec_DPoDframework}

To mitigate APTBM reconstruction degradation while reducing the online learning overhead of the DPoD scheme, this paper proposes a transfer learning-enabled distortion compensation framework. This section presents the proposed DPoD framework, including the APTBM-based frame structure, constraint-guided offline pretraining, and few-shot online adaptation.

\subsection{APTBM-Based Frame Structure Design}
\label{SubSec_FrameDesign}
To support the proposed structure-aware DPoD framework, a dedicated frame structure is designed based on APTBM and the temporal behavior of PA nonlinearity. Unlike fast-varying channel impairments, PA nonlinearity evolves slowly with operating conditions such as temperature and input power back-off \cite{SalmanGuvensen2021_QAMDetector_NonlinearDPoD_FDEBank}. This time-scale disparity motivates a frame design that separates slow-time model adaptation from fast-time (FT) data transmission, achieving nonlinear compensation model offline generalization learning and online adaptation.

\begin{figure}[tp]%
\centering
\includegraphics[width=3.4in,trim=6mm 8mm 8mm 6mm, clip]{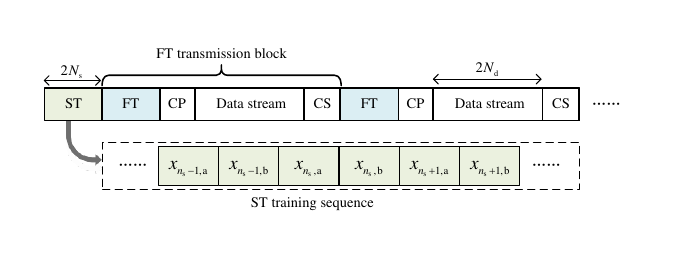}
\caption{APTBM-based frame structure with ST training sequence and FT transmission Blocks.}
\label{fig_APTBM_frame}
\vspace{-6 pt}
\end{figure}

As shown in Fig. \ref{fig_APTBM_frame}, each transmission frame consists of two types: ST training sequences and FT transmission blocks, both constructed using the APTBM modulation format. $N_\mathrm{s}$ and $N_\mathrm{d}$ denote the numbers of APTBM symbol-blocks in the ST training sequence and the data transmission sequence, respectively. Specifically, the ST training sequence is introduced for online adaptation of the PA nonlinearity compensation at the receiver. It is constructed using APTBM symbol-blocks $\mathbf{x}_{n_\mathrm{s}} = [x_{n_\mathrm{s},\mathrm{a}}, x_{n_\mathrm{s},\mathrm{b}}]$, where $n_\mathrm{s}=1,2,\ldots,N_\mathrm{s}$. Since the ST sequence is dedicated to model adaptation rather than synchronization or channel estimation, it is transmitted only when the PA operating condition changes, resulting in negligible overhead \cite{SalmanGuvensen2021_QAMDetector_NonlinearDPoD_FDEBank}.

Each FT transmission block consists of an FT training sequence, a cyclic prefix (CP), an APTBM-modulated data stream, and a cyclic suffix (CS). The FT training sequence constructed using a Golay complementary sequence pair that is used for time synchronization and channel estimation. The CP and CS are appended to facilitate block-based processing and to improve the stability of frequency-domain equalization and nonlinear compensation. In contrast to the ST training sequence, FT transmission blocks are repeatedly transmitted to support fast-time channel tracking and continuous data transmission.

\subsection{Offline Weakly Supervised Model Learning}
\label{SubSec_offlineLearning}

In practical uplink and satellite communication scenarios, DPoD schemes often lack sufficient high-quality transmitter-side signals for model training, limiting their deployment and high-PAE operation. To address this issue, we introduce an offline weakly supervised inverse model learning stage, which exploits abundant received data and the computational resources available at the receiver to provide reliable model initialization for subsequent online few-shot adaptation, without requiring ideal transmitter-side training sequences.

\begin{figure*}[tph]%
\centering
\includegraphics[width=6.8in,trim=6mm 6mm 6mm 6mm, clip]{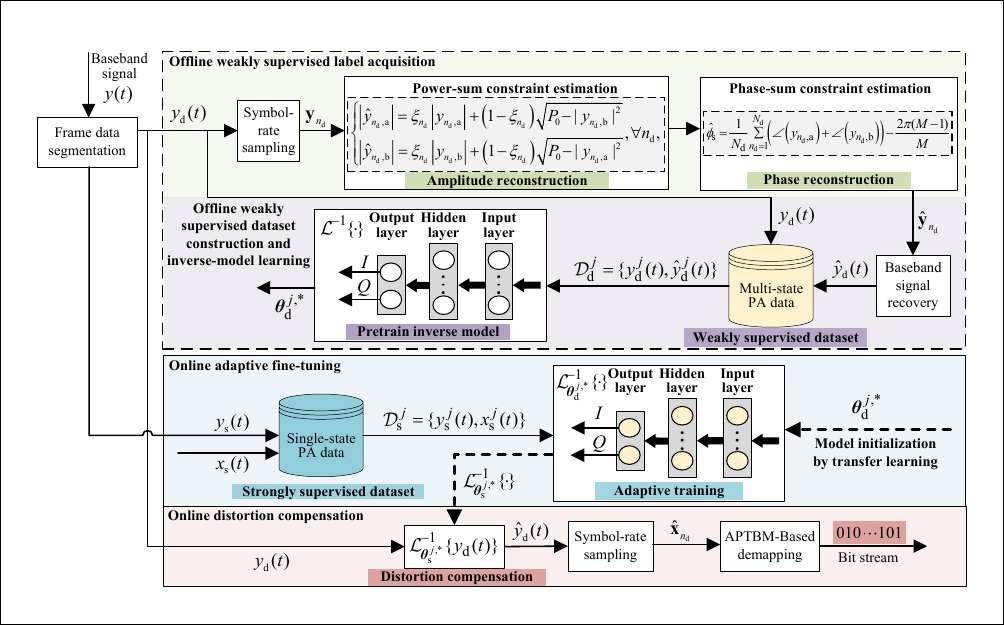}
\caption{Block diagram of the proposed APTBM-based DPoD framework with offline weakly supervised model pretraining and online few-shot adaptation.}
\label{fig_DPoDflowchart}
\vspace{-6 pt}
\end{figure*}

As illustrated in Fig. \ref{fig_DPoDflowchart}, the offline stage consists of two parts: weakly supervised label construction and inverse-model learning. The received discrete-time baseband signal $y(t)$, which contains the data stream and the ST training sequence, is obtained following RF front-end processing and channel equalization. In accordance with the predefined frame structure, we first extract the data-stream component $y_\mathrm{d}(t)$ and then perform symbol-rate sampling to yield the distorted data symbol-block sequence. The $n_\mathrm{d}$-th data symbol-block is denoted as $\mathbf{y}_{n_\mathrm{d}} = [y_{n_{\mathrm{d}},\mathrm{a}}, y_{n_{\mathrm{d}},\mathrm{b}}]$, where $n_{\mathrm{d}}=1,\ldots,N_{\mathrm{d}}$. 

As discussed in Section \ref{SubSec_APTBM}, each symbol-block satisfies the power-sum constraint given in (\ref{Eq_symbolBlock_PowerSum}). Due to AM–AM distortion, higher-amplitude symbols experience stronger compression at the PA output, whereas lower-amplitude symbols are comparatively less affected. Thus, within a symbol-block, higher-amplitude symbols are more likely to operate in the PA compression region, whereas lower-amplitude symbols tend to remain in the near-linear region \cite{Fan2026_APFBM}. This complementary behavior enables block-wise amplitude reconstruction. Specifically, the amplitudes of the reconstructed symbol-block $\hat{\mathbf{y}}_{n_\mathrm{d}} = [\hat{y}_{n_{\mathrm{d}},\mathrm{a}}, \hat{y}_{n_{\mathrm{d}},\mathrm{b}}]$ are obtained by cross-estimation compensation and given by
\begin{equation}
\label{Eq_Rx_PowerSum_AmpEst}
\begin{cases}
\left|\hat{y}_{n_\mathrm{d},\mathrm{a}}\right|=\xi_{n_\mathrm{d}}\left|y_{n_\mathrm{d},\mathrm{a}}\right|+\left(1-\xi_{n_\mathrm{d}}\right)\sqrt{P_0-\mid y_{n_\mathrm{d},\mathrm{b}}\mid^2}, \\
\left|\hat{y}_{n_\mathrm{d},\mathrm{b}}\right|=\xi_{n_\mathrm{d}}\left|y_{n_\mathrm{d},\mathrm{b}}\right|+\left(1-\xi_{n_\mathrm{d}}\right)\sqrt{P_0-\mid y_{n_\mathrm{d},\mathrm{a}}\mid^2},
\end{cases} 
\end{equation}
where $\xi_{n_\mathrm{d}} \in (0,1)$ is an adaptive weight adjusted according to the power difference within each symbol-block. It is defined by a modified sigmoid function as
\begin{equation}
\label{Eq_Rx_Powerdiff_xi}
\xi_{n_\mathrm{d}}=\frac{1}{1+e^{\tan(0.5\pi P_{{n_\mathrm{d}},\mathrm{e}})}},
\end{equation}
where the block-wise power difference $P_{{n_\mathrm{d}},\mathrm{e}}$ is given by
\begin{equation}
\label{Eq_Rx_Powerdiff}
P_{{n_\mathrm{d}},\mathrm{e}}=\frac{|y_{{n_\mathrm{d}},\mathrm{a}}|^2-|y_{{n_\mathrm{d}},\mathrm{b}}|^2}{|y_{{n_\mathrm{d}},\mathrm{a}}|^2+|y_{{n_\mathrm{d}},\mathrm{b}}|^2}.
\end{equation}
The complementary estimation is emphasized for strongly imbalanced blocks, whereas the received amplitudes dominate when the two components have similar powers.

Furthermore, under PA nonlinear operation, AM–PM distortion arises from amplitude-dependent phase rotation. The two symbols within each symbol-block are temporally adjacent and subject to a power-sum constraint, which results in strongly correlated amplitudes. Consequently, the PA-induced phase distortion manifests primarily as a correlated common phase rotation across the symbol-block, accompanied by a residual power-dependent phase shift within the block \cite{Fan2025_APTBM}. Accordingly, the phase of a received symbol-block can be approximated as
\begin{equation}
\label{Eq_phaseShift_Compose}
\begin{cases}
\angle\left(y_{n_\mathrm{d},\mathrm{a}}\right)=\angle\left(x_{n_\mathrm{d},\mathrm{a}}\right)+\frac{1}{2}\phi_\mathrm{s}+\vartheta_{n_\mathrm{d},\mathrm{a}}, \\
\angle\left(y_{n_\mathrm{d},\mathrm{b}}\right)=\angle\left(x_{n_\mathrm{d},\mathrm{b}}\right)+\frac{1}{2}\phi_\mathrm{s}+\vartheta_{n_\mathrm{d},\mathrm{b}}, & 
\end{cases}
\end{equation}
where $\phi_\mathrm{s}$ represents the block-level common phase rotation term, while $\vartheta_{n_\mathrm{d},\mathrm{a}}$ and $\vartheta_{n_\mathrm{d},\mathrm{b}}$ denote the residual phase shifts. The practical phase shift of intra-block symbols fluctuates around the average common phase shift $(\frac{1}{2}\phi_\mathrm{s})$.

From (\ref{Eq_four_relativeState}), the phase-sum of an ideal APTBM symbol-block belongs to a known PSK alphabet set \cite{Fan2025_APTBM}. Accordingly, the common phase rotation induced by PA nonlinearity can be estimated by evaluating the statistical shift of the phase-sum between the transmitted and received symbol-blocks, given by
\begin{equation}
\label{Eq_phaseShift_mean}
\begin{aligned}
\hat{\phi}_\mathrm{s} 
&= \mathbb{E}\{\angle\left(y_{n_\mathrm{d},\mathrm{a}}\right) + \angle\left(y_{n_\mathrm{d},\mathrm{b}}\right)\}
   - \mathbb{E}\{\angle\left(x_{n_\mathrm{d},\mathrm{a}}\right) + \angle\left(x_{n_\mathrm{d},\mathrm{b}}\right)\} \\
&= \frac{1}{N_{\mathrm{d}}} \sum_{n_\mathrm{d}=1}^{N_\mathrm{d}} \left( \angle\left(y_{n_\mathrm{d},\mathrm{a}}\right) + \angle\left(y_{n_\mathrm{d},\mathrm{b}}\right)\right)
   - \frac{2\pi (M-1)}{M}.
\end{aligned}
\end{equation}
Subsequently, to account for residual phase distortion, the power-dependent adaptive weighting factor $\xi_{n_\mathrm{d}}$ in (\ref{Eq_Rx_Powerdiff_xi}) is introduced to asymmetrically allocate the common phase shift within each symbol-block. Thus, the reconstructed phase is obtained by removing the estimated nonlinear phase shift from the received phase, yielding
\begin{equation}
\label{Eq_phaseShift_Est}
\begin{cases}
\angle\left(\hat{y}_{n_\mathrm{d},\mathrm{a}}\right) = \angle\left(y_{n_\mathrm{d},\mathrm{a}}\right) - \left(\frac{3}{4}-\frac{1}{2}\xi_{n_\mathrm{d}}\right)\hat{\phi}_\mathrm{s}, \\
\angle\left(\hat{y}_{n_\mathrm{d},\mathrm{b}}\right)=  \angle\left(y_{n_\mathrm{d},\mathrm{b}}\right) - \left(\frac{1}{4}+\frac{1}{2}\xi_{n_\mathrm{d}}\right)\hat{\phi}_\mathrm{s}.  
\end{cases}
\end{equation}

In summary, APTBM symbol-block constraints and PA distortion statistical characteristics enable feasible receiver-side amplitude and phase reconstruction without ideal training sequences. Although these block constraints do not fully characterize nonlinear and memory-dependent PA behavior, they yield physically plausible weak labels that restrict the inverse-model search space and provide robust initialization.

Using the signals before and after reconstruction, a weakly supervised dataset can be constructed for PA inverse-model training. The reconstructed symbol-block sequence is upsampled and pulse-shaped to reconstruct the continuous-time baseband waveform $\hat{y}_{\mathrm{d}}(t)$, which is required because PA distortion is introduced after pulse shaping. The above reconstruction process can be applied to the received data collected under different operating states of the PA. Aggregating these results yields a weakly supervised dataset
\begin{equation}
\label{Eq_weakDataset}
\mathcal{D}_{\mathrm{d}}^{j}=\{y_\mathrm{d}^{j}(t),\hat{y}_\mathrm{d}^{j}(t)\}, \forall t,
\end{equation}
where $j$ denotes the different operating states of the PA. Based on the weakly supervised dataset $\mathcal{D}_\mathrm{d}^{j}$ and the nonlinear nature of the PA, the inverse-model function $\mathcal{L}^{-1}_{\boldsymbol{\theta}_\mathrm{d}^j}\{\cdot\}$ is defined to minimize the error between the signals before and after reconstruction. Thus, the offline inverse-model learning problem is formulated as

\begin{equation}
\label{Eq_offline_pretrainingProblem}
\boldsymbol{\theta}_{\mathrm{d}}^{j,*}
=
\underset{\boldsymbol{\theta}_\mathrm{d}^{j}}{\arg\min}\,
\frac{1}{\left|\mathcal{D}_{\mathrm{d}}^{j}\right|}
\sum_{t\in\mathcal{D}_{\mathrm{d}}^{j}}
\left\|
\mathcal{L}^{-1}_{\boldsymbol{\theta}_\mathrm{d}^{j}}
\left\{y_\mathrm{d}^{j}(t)\right\}
-
\hat{y}_\mathrm{d}^{j}(t)
\right\|_2^2 .
\end{equation}

The objective of model learning is to determine a set of model parameters $\boldsymbol{\theta}_\mathrm{d}^{j,*}$ such that the model maps the distorted signal to its constraint-reconstructed reference. These offline-derived inverse models are not intended to precisely characterize PA nonlinearity. Instead, they provide effective initialization for subsequent online adaptive learning. 

\subsection{Online Few-Shot Adaptation and Distortion Compensation}
\label{SubSec_onlineFT}
In conventional DPoD schemes \cite{PihlajasaloEtAl2023_DeepLearningOFDMReceiversPowerEfficiency}, online inverse-model learning suffers from training sequence overhead and small-sample overfitting, which can degrade convergence stability and compensation reliability. To address these limitations, transfer learning bridges offline pretraining and online adaptation, enabling efficient inverse-model updating with reduced online training overhead while maintaining high compensation accuracy.

By transferring model knowledge obtained from offline learning to initialize the online inverse model for the current PA operating state, the model can be efficiently fine-tuned using ST training sequences, enabling rapid adaptation to dynamic scenario variations. As discussed in Section \ref{SubSec_FrameDesign}, PA nonlinearity evolves on a slow time scale. Thus, online adaptation is performed using the ST training sequences sparsely inserted in the frame. As illustrated in Fig. \ref{fig_DPoDflowchart}, the baseband signal $y_{\mathrm{s}}(t)$ corresponding to the ST training sequence is extracted via frame segmentation. Since the ST training sequence is known at the receiver, a strongly supervised dataset can be constructed from the ST symbols, given by
\begin{equation}
\label{Eq_strongDataset}
\mathcal{D}_{\mathrm{s}}^{j}=\{y_\mathrm{s}^{j}(t),x_\mathrm{s}^{j}(t)\}, \forall t.
\end{equation}

To ensure structural consistency between offline and online inverse model functions, the online inverse model learning problem is formulated as 
\begin{equation}
\label{Eq_online_adaptationProblem}
\begin{aligned}
\boldsymbol{\theta}_{\mathrm{s}}^{j,*}
=
\underset{\boldsymbol{\theta}_{\mathrm{s}}^{j}}{\arg\min}
\Bigg[
&\frac{1}{\left|\mathcal{D}_{\mathrm{s}}^{j}\right|}
\sum_{t\in\mathcal{D}_{\mathrm{s}}^{j}}
\left\|
\mathcal{L}^{-1}_{\boldsymbol{\theta}_{\mathrm{s}}^{j}}
\left\{y_{\mathrm{s}}^{j}(t)\right\}
-
x_{\mathrm{s}}^{j}(t)
\right\|_2^2
\\
&\quad+
\lambda_d
\left\|
\boldsymbol{\theta}_{\mathrm{s}}^{j}
-
\boldsymbol{\theta}_{\mathrm{d}}^{j,*}
\right\|_2^2
\Bigg].
\end{aligned}
\end{equation}
where $\boldsymbol{\theta}_\mathrm{d}^{j,*}$ and $\boldsymbol{\theta}_\mathrm{s}^{j,*}$ denote the offline-pretrained and online-adapted optimal inverse-model parameters for the $j$-th PA operating state, respectively. $\lambda_d > 0$ is a smoothing coefficient that controls the update rate of model parameters during online adaptation. A smaller $\lambda_d$ improves tracking capability under rapidly varying nonlinear conditions, whereas a larger $\lambda_d$ is suitable when the PA characteristics vary slowly.

\begin{figure}[tp]%
\centering
\includegraphics[width=3.4in,trim=8mm 6mm 6mm 6mm, clip]{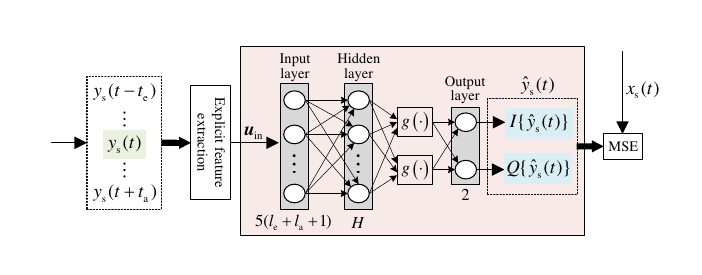}
\caption{Lightweight PA inverse-model learning in the online stage.}
\label{fig_LightweightPAInverseModel}
\vspace{-6 pt}
\end{figure}

After the adaptive learning of the online model, the updated PA inverse model is applied to compensate for distortion in the data-stream,
\begin{equation}
\label{Eq_Rx_PAInverseMapping}
\hat{x}_\mathrm{d}(t)=\mathcal{L}^{-1}_{\boldsymbol{\theta}_\mathrm{s}^{j,*}}\{y_\mathrm{d}(t)\}.
\end{equation}
Subsequently, the compensated APTBM symbol-blocks are recovered via symbol-rate sampling. These symbols are then demapped into the transmitted bit stream by channel decoding.

To improve the representation capability for PA nonlinear behavior while reducing the computational overhead of online compensation, a lightweight neural network model is developed to learn the inverse PA distortion characteristics by explicit feature extraction (EFE) from the received signal. Notably, each signal sample is affected not only by PA nonlinearity and memory effects but also by the coupled memory introduced by pulse shaping and filtering. Therefore, unlike prior models that consider only postcursor samples \cite{WangEtAl2019_ARVTDNN,Rawat2012_RVFTDNN}, the proposed model jointly exploits both precursor and postcursor samples to characterize their impact on the current sample. EFE is further employed to represent the PA nonlinear behavior efficiently through low-dimensional feature vector mappings. Taking ST training sequence signal samples as an example, the feature vector $\boldsymbol{\mu}_{\mathrm{in}}$ is given by
\begin{equation}
\label{Eq_featureExtraction}
\begin{aligned}
\boldsymbol{\mu}_{\mathrm{in}}(t) = 
\big[ &
I\{y_{\mathrm{s}}(t-l_\mathrm{e})\}, \ldots, I\{y_{\mathrm{s}}(t)\}, \ldots, I\{y_{\mathrm{s}}(t+l_\mathrm{a})\}, \\
& Q\{y_{\mathrm{s}}(t-l_\mathrm{e})\}, \ldots, Q\{y_{\mathrm{s}}(t)\}, \ldots, Q\{y_{\mathrm{s}}(t+l_\mathrm{a})\}, \\
& |y_{\mathrm{s}}(t-l_\mathrm{e})|, \ldots, |y_{\mathrm{s}}(t)|, \ldots, |y_{\mathrm{s}}(t+l_\mathrm{a})|, \\
& |y_{\mathrm{s}}(t-l_\mathrm{e})|^{2}, \ldots, |y_{\mathrm{s}}(t)|^{2}, \ldots, |y_{\mathrm{s}}(t+l_\mathrm{a})|^{2}, \\
& |y_{\mathrm{s}}(t-l_\mathrm{e})|^{3}, \ldots, |y_{\mathrm{s}}(t)|^{3}, \ldots, |y_{\mathrm{s}}(t+l_\mathrm{a})|^{3}
\big]^\mathrm{T},
\end{aligned}
\end{equation}
where $l_\mathrm{e}$ and $l_\mathrm{a}$ represent the numbers of postcursor and precursor samples, respectively. The constructed feature vector incorporates the real and imaginary components of the distorted signal, as well as multiple envelope-related terms, enabling effective characterization of PA nonlinearity and memory effects \cite{WangEtAl2019_ARVTDNN}.

Fig. \ref{fig_LightweightPAInverseModel} illustrates the construction and learning process of the lightweight PA inverse model. After feature extraction from the received signal $y_{\mathrm{s}}(t)$, a single hidden-layer feedforward neural network is employed to model the inverse mapping of PA. The model input is the feature vector $\boldsymbol{\mu}_{\mathrm{in}}(t)$, whose dimension $5(l_{\mathrm{e}}+l_{\mathrm{a}}+1)$ is determined by the feature design and the memory depth. The hidden layer consists of $H$ neurons to capture nonlinear feature interactions, and a nonlinear activation function $g(\cdot)$ is applied to enhance the representation capability. The model outputs a two-dimensional real-valued vector representing the in-phase and quadrature components of the distortion-compensated signal $\hat{y}_{\mathrm{s}}(t)$, denoted as $\boldsymbol{\mu}_{\mathrm{out}}(t) = \big[ I\{\hat{y}_{\mathrm{s}}(t)\},\, Q\{\hat{y}_{\mathrm{s}}(t)\} \big]^{\mathrm{T}}$. The forward computation of the inverse model is given by
\begin{equation}
\label{Eq_NNmodel}
\boldsymbol{\mu}_{\mathrm{out}}(t)
= \boldsymbol{W}_2 \, g\!\left(\boldsymbol{W}_1 \boldsymbol{\mu}_{\mathrm{in}}(t) + \boldsymbol{b}_1 \right)
+ \boldsymbol{b}_2 ,
\end{equation}
where $\boldsymbol{W}_1 \in \mathbb{R}^{H \times 5(l_{\mathrm{e}}+l_{\mathrm{a}}+1)}$ and $\boldsymbol{b}_1 \in \mathbb{R}^{H \times 1}$ represent the weight matrix and bias vector of the hidden layer, respectively, and $\boldsymbol{W}_2 \in \mathbb{R}^{2 \times H}$ and $\boldsymbol{b}_2 \in \mathbb{R}^{2 \times 1}$ correspond to the output layer. The total number of trainable parameters in the model is given by $N_{\mathrm{par}} = H \big(5(l_{\mathrm{e}}+l_{\mathrm{a}}+1) + 3 \big) + 2$. 

\begin{figure}[tp]%
\centering
\includegraphics[width=3.4in,trim=6mm 6mm 4mm 6mm, clip]{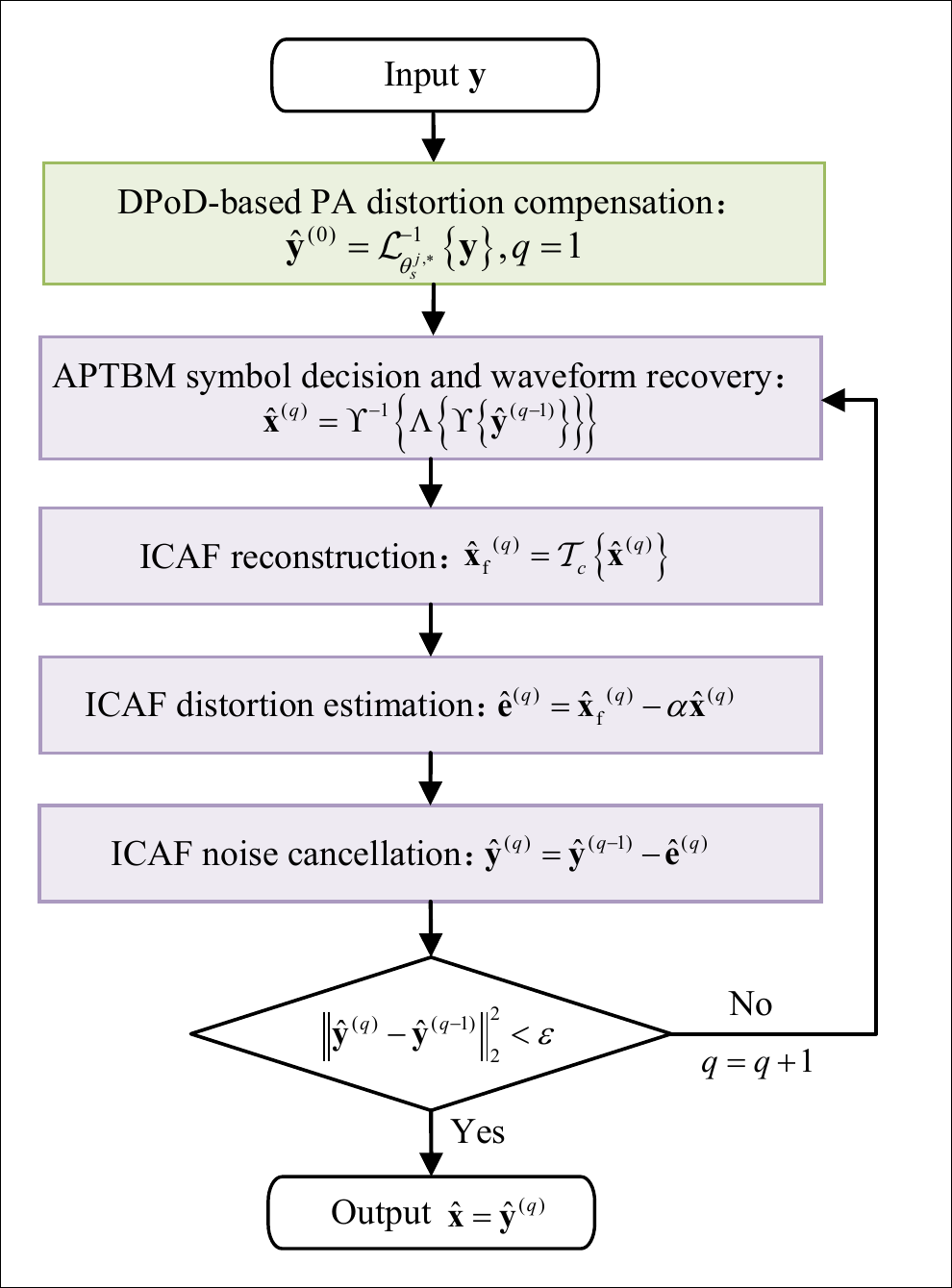}
\caption{Cascaded distortion compensation flowchart for PA and ICAF.}
\label{fig_CascadeCompensationScheme}
\vspace{-6 pt}
\end{figure}

Additionally, the mean squared error (MSE) is adopted as the loss function between the distorted signal samples and the ideal signal samples. The model parameters are optimized through backpropagation using the Adam optimizer, and the hyperbolic tangent sigmoid (tansig) function is employed as the nonlinear activation function $g(\cdot)$ \cite{WangEtAl2019_ARVTDNN}. 

\section{Compensation for Nonlinear Distortion Induced by PA and ICAF}
\label{Sec_cascadedCompensation}

PA operation near saturation can induce severe spectral regrowth, causing OOB emissions that interfere with adjacent RF channels. Therefore, ACLR compliance must be addressed together with in-band nonlinear distortion. In the proposed transceiver framework, ICAF and SDPD are employed at the transmitter to reduce PAPR and suppress OOB emissions. However, the cascaded ICAF operation and residual PA nonlinearity introduce coupled in-band distortion at the receiver. Building upon the DPoD scheme proposed in Section~\ref{Sec_DPoDframework}, a receiver-side cascaded DPoD and CNC compensation scheme is developed. 

As illustrated in Fig. \ref{fig_CascadeCompensationScheme}, for the equalized received baseband sequence $\mathbf{y}$, the dominant impairments are assumed to be the ICAF-induced distortion and the residual PA nonlinear distortion, while other channel impairments are compensated. Following the inverse order of the transmitter-side ICAF and PA processing, the PA-induced distortion is first mitigated. During the online stage, the ST training sequence $\mathbf{x}_{\mathrm{s}}$ is processed by the same ICAF operator as that used at the transmitter, yielding $\bar{\mathbf{x}}_{\mathrm{s}}=\mathcal{T}_{c}\{\mathbf{x}_{\mathrm{s}}\}$. Accordingly, the strongly supervised dataset for online adaptation is constructed as $\{\mathbf{y}_{\mathrm{s}},\bar{\mathbf{x}}_{\mathrm{s}}\}$, where the target signal contains the deterministic ICAF distortion but excludes the PA nonlinear distortion. Similarly, in the offline weakly supervised stage, the reconstructed reference in \eqref{Eq_weakDataset} is replaced by its ICAF-processed counterpart $\bar{\mathbf{y}}_{\mathrm{d}}^{j}=\mathcal{T}_{c}\{\hat{\mathbf{y}}_{\mathrm{d}}^{j}\}$.
Thus, the DPoD model is trained to map the received distorted waveform to a ICAF-dominated reference, rather than directly to the ideal undistorted waveform. Although the ICAF distortion remains in the training target, the network learns the residual difference between the received signal and the ICAF-processed reference, thereby enabling effective characterization and compensation of the PA nonlinear residual. 

After PA distortion compensation, the remaining distortion in $\hat{\mathbf{y}}^{(0)}$ is mainly caused by ICAF. Thus, a decision-directed CNC procedure is introduced. At the $q$-th CNC iteration, the current waveform is first detected in the APTBM symbol domain and then regenerated into a clean baseband waveform as
\begin{equation}
\label{Eq_CNCWaveformRecovery}
\hat{\mathbf{x}}^{(q)}=
\Upsilon^{-1}\left\{\Lambda\left\{\Upsilon
\left\{\hat{\mathbf{y}}^{(q-1)}\right\}\right\}\right\},
\end{equation}
where $\Upsilon\{\cdot\}$ denotes matched filtering and symbol-rate sampling, $\Lambda\{\cdot\}$ denotes APTBM symbol-block decision and remapping, and $\Upsilon^{-1}\{\cdot\}$ represents pulse-shaped waveform recovery. Using the known clipping threshold and filtering response, the ICAF output corresponding to $\hat{\mathbf{x}}^{(q)}$ is reconstructed by $\hat{\mathbf{x}}_{f}^{(q)}=\mathcal{T}_{c}\left\{\hat{\mathbf{x}}^{(q)}\right\}$.
The ICAF-induced distortion is then estimated as
\begin{equation}
\label{Eq_CAFNoise_est}
\hat{\mathbf{e}}^{(q)}=
\hat{\mathbf{x}}_{f}^{(q)}-\alpha\hat{\mathbf{x}}^{(q)},
\end{equation}
where $\alpha=0.99$ is a scaling factor used to compensate for the gain mismatch between the reconstructed and received waveforms. The estimated ICAF distortion is cancelled from the current waveform according to
\begin{equation}
\label{Eq_CAFNoise_elim}
\hat{\mathbf{y}}^{(q)}=
\hat{\mathbf{y}}^{(q-1)}-\hat{\mathbf{e}}^{(q)}.
\end{equation}
The iteration stops when
\begin{equation}
\left\|\hat{\mathbf{y}}^{(q)}-
\hat{\mathbf{y}}^{(q-1)}\right\|_{2}^{2}<\epsilon,
\end{equation}
or when the preset maximum number of iterations is reached. $\epsilon$ is set to $10^{-3}$, which represents the convergence threshold. Finally, the compensated waveform is obtained as $\hat{\mathbf{x}}=\hat{\mathbf{y}}^{(q)}$. 

The online complexity of the proposed cascaded scheme is dominated by DPoD and iterative CNC processing. For $N_\mathrm{samp}$ received samples and $H$ hidden neurons, the DPoD forward pass requires $\mathcal{O}(N_\mathrm{samp}H)$ operations. In each CNC iteration, reconstructing $K$ ICAF rounds involves frequency-domain filtering via FFT/IFFT operations, yielding a complexity of $\mathcal{O}[KN(\log N_\mathrm{samp}+1)]$. Accordingly, the online complexity of the cascaded scheme is $\mathcal{O}\!\left[N_\mathrm{samp}H+qKN_\mathrm{samp}(\log N_\mathrm{samp}+1)\right]$.

The proposed cascaded structure separates the PA- and ICAF-induced distortions into two sequential compensation stages. The DPoD suppresses the residual PA nonlinear distortion by learning the mapping from the received waveform to the ICAF-dominated reference, while the CNC exploits the deterministic ICAF model to iteratively estimate and remove the residual clipping-filtering noise. This design avoids explicitly learning the ICAF-PA cascaded inverse process and provides a receiver-side solution with superior compensation performance for ACLR-constrained high-efficiency PA operation.

\section{Experimental Results and Analysis}
\label{Sec_Exper_Anal}

This section evaluates ACLR, IBO, PAE, EVM, and BER through simulation and measurement platforms. The proposed transceiver-cooperative framework, including the receiver-side DPoD and cascaded compensation schemes, is assessed under different PA models.

\subsection{Experimental Setup and Performance Metrics}
\label{SubSec_SetupAndMetrics}

To evaluate PA-induced spectral regrowth, the ACLR is measured
according to the specification
\cite{3GPP_TS381015}. Let $U_{\mathrm{out}}(f)$ denote the power spectral density of the PA output signal. The ACLR is defined as
\begin{equation}
\mathrm{ACLR}
=
\min_{u\in\{-1,+1\}}
10\log_{10}\frac{
\displaystyle\int_{\mathcal{B}_{0}}
U_{\mathrm{out}}(f)\,\mathrm{d}f
}{
\displaystyle\int_{\mathcal{B}_{u}}
U_{\mathrm{out}}(f)\,\mathrm{d}f
},
\label{Eq_ACLR}
\end{equation}
where $\mathcal{B}_{0}$ and $\mathcal{B}_{u}$ denote the measurement bands of the assigned channel and the lower or upper adjacent channel, respectively. Following the 3GPP specifications for non-terrestrial network (NTN) scenarios \cite{3GPP_TS381015}, the channel bandwidth and occupied bandwidth are set to 100 MHz and 98.310 MHz, respectively. The UE ACLR requirements of 30 dBc and 45 dBc are adopted as the spectral compliance constraints. 

\begin{figure}[tp]
\centering
\includegraphics[width=3.4in, trim=10mm 12mm 8mm 12mm,clip]{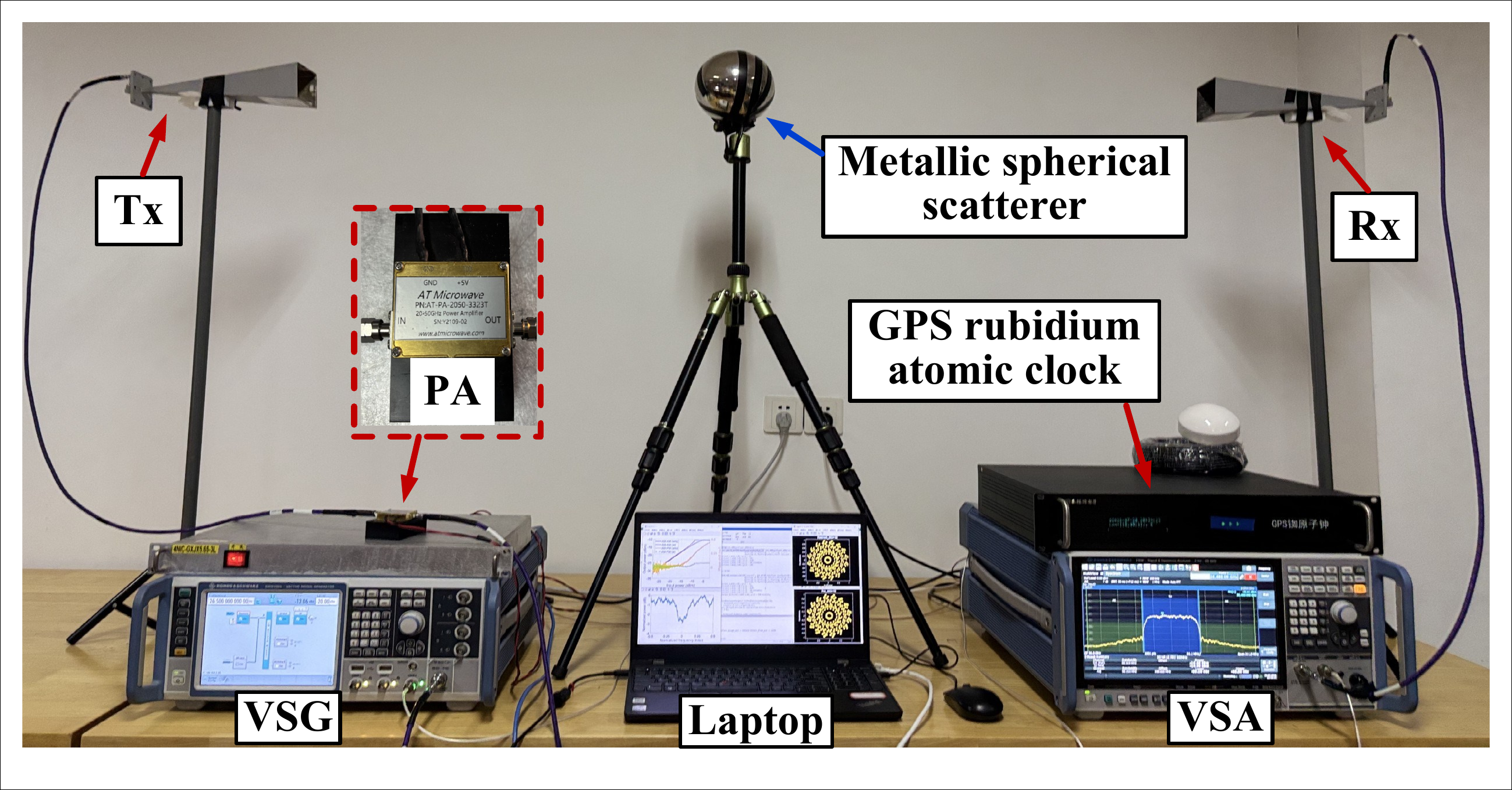}
\caption{Measurement experiment platform and scenario.}
\label{fig_measurementSetup_Scenario}
\vspace{-6 pt}
\end{figure}

The PA operating condition and transmitter-side efficiency gain are evaluated using the IBO and PAE. IBO quantifies the back-off of the PA operating input power \(P_{\mathrm{in}}\) from the \(P_{\mathrm{in,sat}}\), and is defined as
\begin{equation}
\label{Eq_IBO}
\mathrm{IBO}=10\log_{10}\left(\frac{P_{\mathrm{in,sat}}}{P_{\mathrm{in}}}\right).
\end{equation}
For Class A power amplifiers, the PAE ($\eta_{\mathrm{PA}}$) is roughly defined as the ratio of the output power \(P_{\mathrm{out}}\) to saturated output power \(P_{\mathrm{out,sat}}\) \cite{Xia2026_TwoStageAPTBM}:
\begin{equation}
\label{Eq_PAE}
\eta_{\mathrm{PA}} \approx \frac{P_{\mathrm{out}}}{2P_{\mathrm{out,sat}}}.
\end{equation}

To evaluate the distortion compensation performance for modulated symbols, EVM and BER are employed to evaluate the accuracy and reliability of information transmission \cite{Fan2025_APTBM,Duan2026_APSBM}. The system incorporates a low-density parity-check (LDPC) code with a length of $648$ and a rate of $5/6$. Pulse shaping and matched filtering are implemented using root-raised cosine (RRC) filters with a roll-off factor of $0.2$ and an oversampling factor of $4$.

In the numerical simulation, the modified Rapp and Saleh models are adopted to characterize memoryless PA nonlinearities \cite{Fan2025_APTBM}, while a generalized memory polynomial (GMP) model is used to capture nonlinear memory effects. The GMP model is obtained from a 28-GHz gallium nitride (GaN) PA reported in a 3GPP specification \cite{NewRadioAccess3GPP_TR_38.803_v14.4.02017}. The $P_{\mathrm{out,sat}}$ for the Rapp, Saleh, and GMP models are -5 dBm, 0 dBm, and -2.87 dBm, respectively. 

\begin{figure}[tp]%
\centering
\subfigure[][]{
    \label{fig_SDPD_AM_PM_all_sim}
    \includegraphics[width=1.6in, trim=8mm 6mm 7mm 12mm, clip]{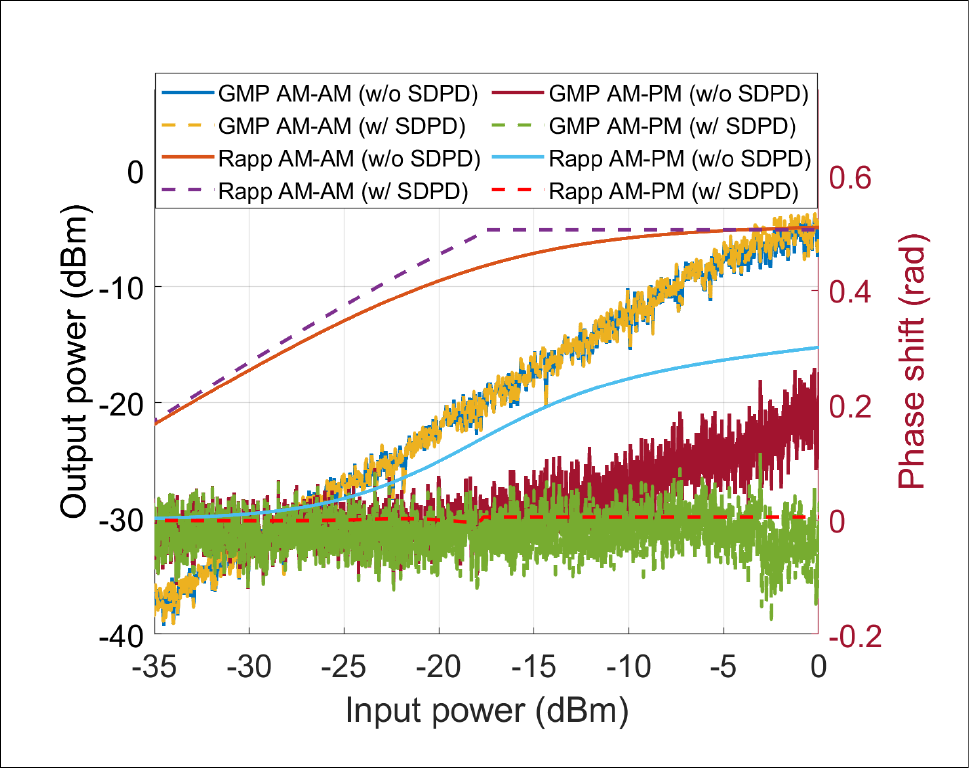}
    }
\subfigure[][]{
    \label{fig_SDPD_test_PA_mea}
    \includegraphics[width=1.6in, trim=8mm 6mm 7mm 12mm, clip]{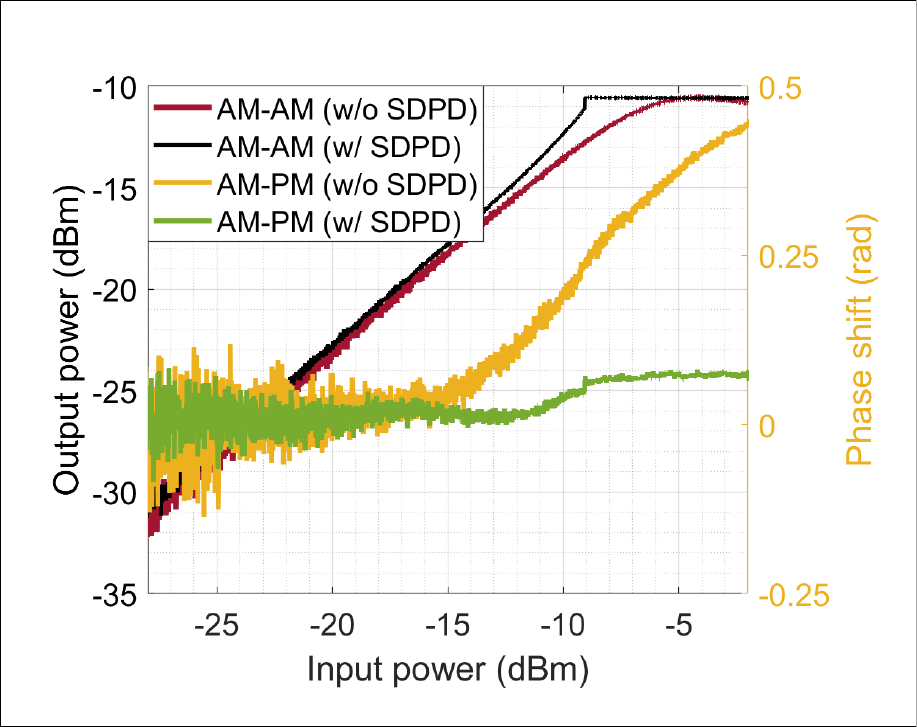}
    }
\subfigure[][]{
    \label{fig_ACLR_PA_Output_mea}
    \includegraphics[width=1.6in, trim=8mm 6mm 12mm 12mm, clip]{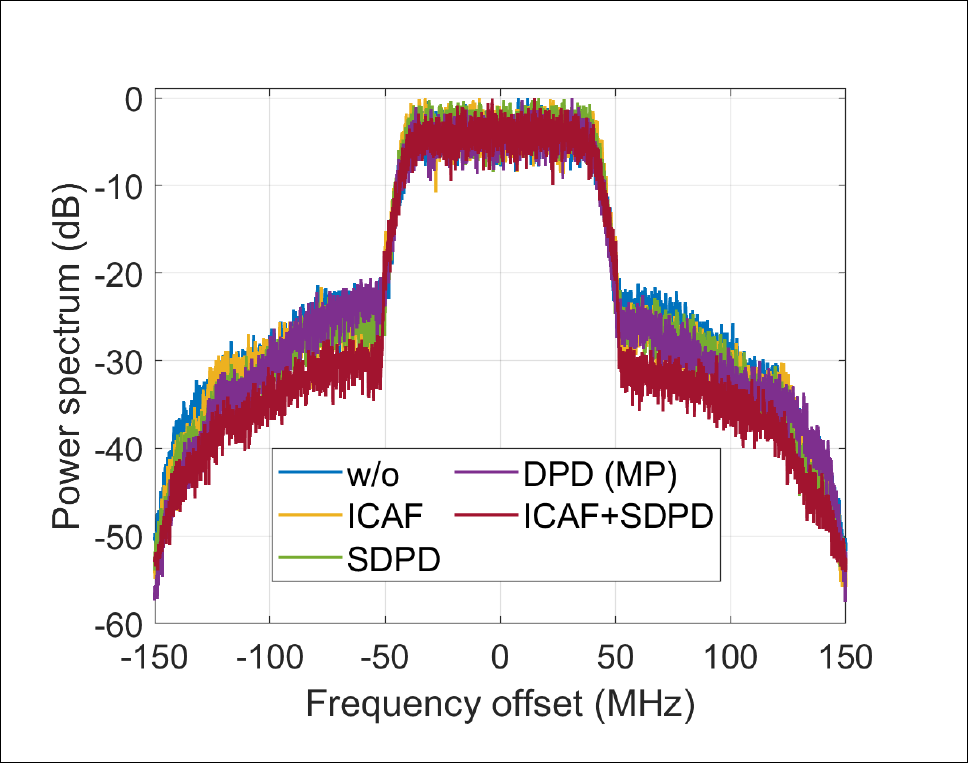}
    }
\subfigure[][]{
    \label{fig_Channel_CFR_mea}
    \includegraphics[width=1.6in, trim=8mm 6mm 12mm 12mm, clip]{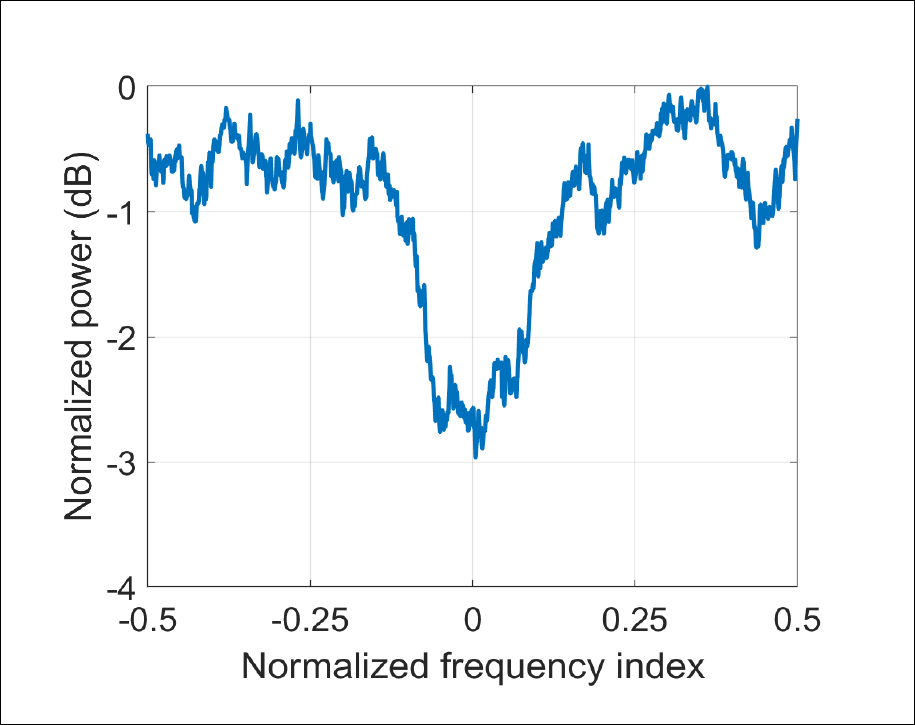}
    }
\caption{SDPD performance and measured channel conditions. (a) and (b) are the simulated and measured PA characteristics before and after SDPD, respectively; (c) The measured spectral emission of the PA output; (d) CFR of the measured scenario.}
\label{fig_SDPD_analysis}
\vspace{-6 pt}
\end{figure}

To validate the applicability and robustness of the proposed schemes in practical scenarios, a measurement platform was developed, as shown in Figs.~\ref{fig_measurementSetup_Scenario}. The platform consists of an R\&S SMW200A vector signal generator (VSG), an R\&S FSW85 vector signal analyzer (VSA), two 25-dBi highly directional horn antennas, and a GPS rubidium atomic clock. All measurements were controlled by a laptop. The experimental measurements utilized a commercial PA operating at 26.5 GHz. A multipath scenario is constructed using a metallic spherical scatterer, with the measured channel frequency response (CFR) shown in Fig. \ref{fig_Channel_CFR_mea}. The Golay complementary sequences enabled channel estimation, while 1-PPS and 10-MHz references provided timing and carrier-frequency synchronization.

\subsection{PA-Nonlinearity Compensation Results}
\label{SubSec_OnlyPAResults}

\begin{figure}[tp]%
\centering
\subfigure[][]{
    \label{fig_Lightweight_Model_Complexity}
    \includegraphics[width=1.6in, trim=8mm 6mm 6mm 12mm, clip]{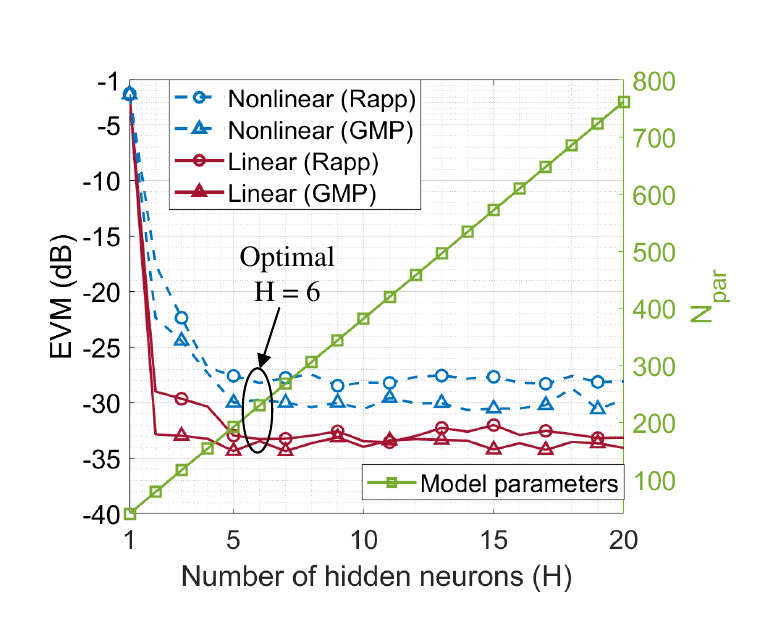}
    }
\subfigure[][]{
    \label{fig_diffPA_InverseModel_EVM}
    \includegraphics[width=1.6in, trim=8mm 6mm 8mm 12mm, clip]{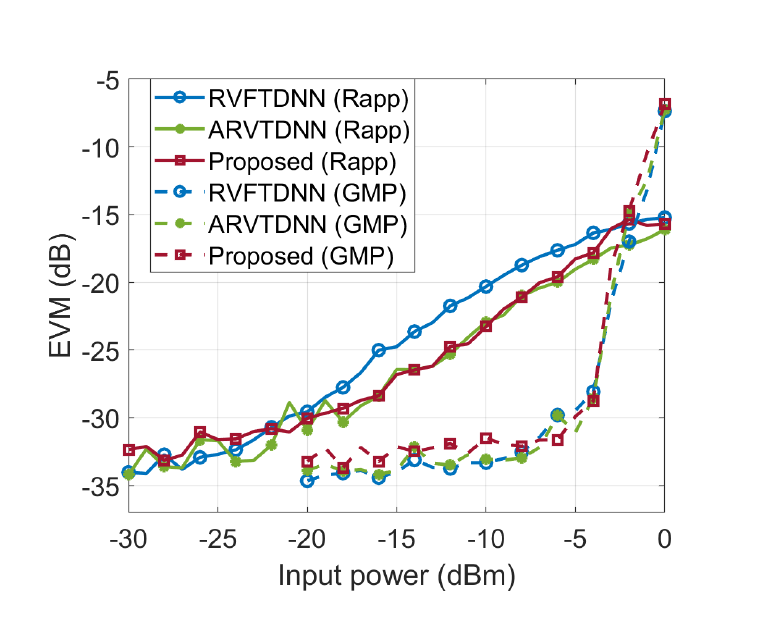}
    }
\caption{The performance analysis of the lightweight PA inverse-model. (a) Model parameters and performance under linear and nonlinear PA operation; (b) PA modeling performance of different network models.}
\label{fig_PAModelComplexityAndPerformance}
\vspace{-6 pt}
\end{figure}

\begin{table}[tp]
\centering
\caption{The computational complexities of different inverse-models}
\label{Tab_PAModelComplexity}
\renewcommand{\arraystretch}{1.3}
\begin{tabular*}{\columnwidth}{@{\extracolsep{\fill}} c p{0.18\columnwidth} c c c c @{}}
\hline
\textbf{Model} & \textbf{Opt. config.} & \textbf{\#Params} & \textbf{\#Mults} & \textbf{\#Adds} \\
\hline
RVFTDNN      & \centering(3,0,5,41,2) & 453 & 410 & 410 \\
ARVTDNN      & \centering(3,0,5,17,2) & 393 & 374 & 374 \\
Proposed model  & \centering(3,3,5,6,2)  & 230 & 222 & 222 \\
\hline
\end{tabular*}
\vspace{0.2em}
\parbox{\columnwidth}{\footnotesize 
\emph{Note:} Opt. config. = (postcursor memory depth, precursor memory depth, input features, hidden neurons, outputs).
\#Params counts trainable parameters. \#Mults and \#Adds are real multiplications and additions per-sample forward pass.}
\end{table}

Using the above experimental setup, the proposed lightweight DPoD model is first evaluated in terms of modeling accuracy and computational complexity. The evaluation considers modified Rapp and GMP behavioral models with 16-APTBM modulation at SNR = $30$ dB, where the online training sequence length is set to $N_\mathrm{s}=1000$. In the offline stage, $140,000$ and $60,000$ symbols are used for training and validation, respectively. Both offline and online models are trained for $100$ epochs with a batch size of $256$, and the learning rate and smoothing factor $\lambda_d$ are set to $0.001$. 

Fig. \ref{fig_Lightweight_Model_Complexity} investigates the impact of hidden-layer size $H$ on EVM performance and model complexity. The linear operating points for the Rapp and GMP models are set to -30 dBm and -20 dBm, while the highly nonlinear operating points are -15 dBm and -4 dBm, respectively. As $H$ increases, the EVM gradually saturates, whereas the number of trainable parameters increases monotonically. Therefore, $H=6$ is selected as the optimal trade-off between compensation accuracy and computational overhead. Additionally, Fig. \ref{fig_diffPA_InverseModel_EVM} and Table \ref{Tab_PAModelComplexity} compare the proposed model with RVFTDNN \cite{Rawat2012_RVFTDNN} and ARVTDNN \cite{WangEtAl2019_ARVTDNN} under linear, nonlinear, and memory-dependent PA conditions. With 1000 iterations for steady-state convergence, the proposed model achieves comparable modeling accuracy while requiring fewer parameters and lower computational complexity. This advantage is attributed to the joint exploitation of precursor and postcursor memory features within a compact architecture, demonstrating its suitability for resource-efficient PA behavioral modeling.

\begin{figure}[tp]%
\centering
\subfigure[][]{
    \label{fig_sim_DPoD_diffInputPower_EVM_Rapp}
    \includegraphics[width=1.6in, trim=8mm 6mm 12mm 12mm, clip]{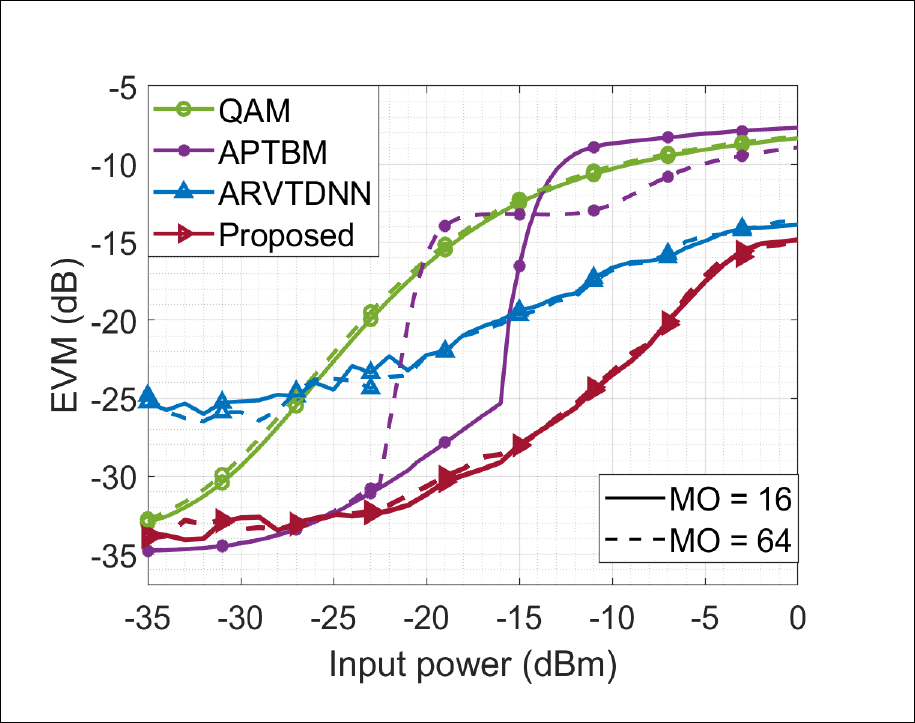}
    }
\subfigure[][]{
    \label{fig_sim_DPoD_diffSNR_BER_Rapp}
    \includegraphics[width=1.6in, trim=8mm 6mm 12mm 11mm, clip]{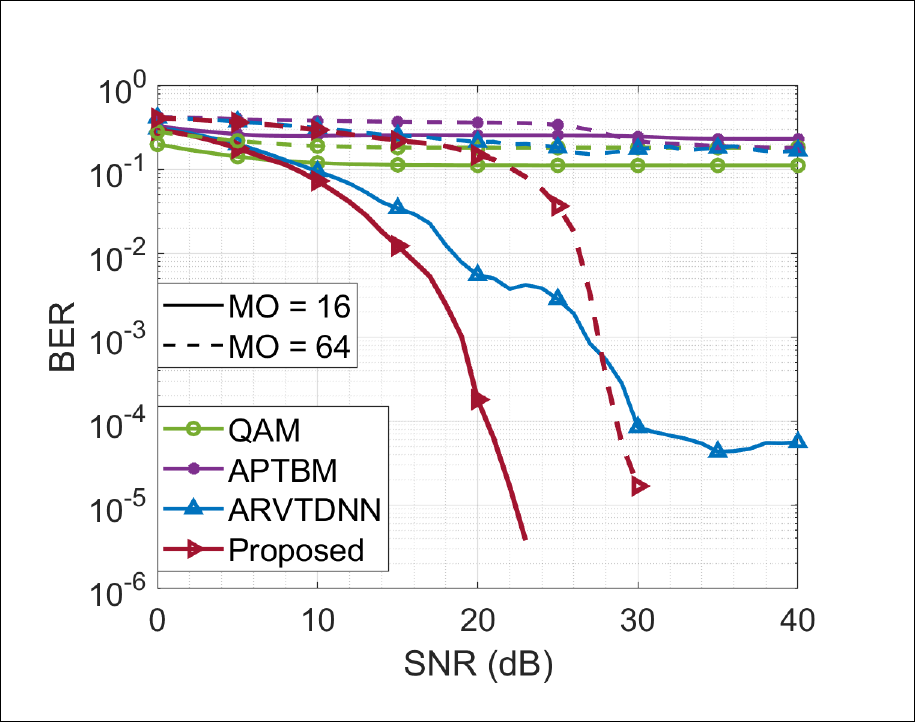}
    }
\subfigure[][]{
    \label{fig_sim_DPoD_diffInputPower_EVM_Saleh}
    \includegraphics[width=1.6in, trim=8mm 6mm 12mm 12mm, clip]{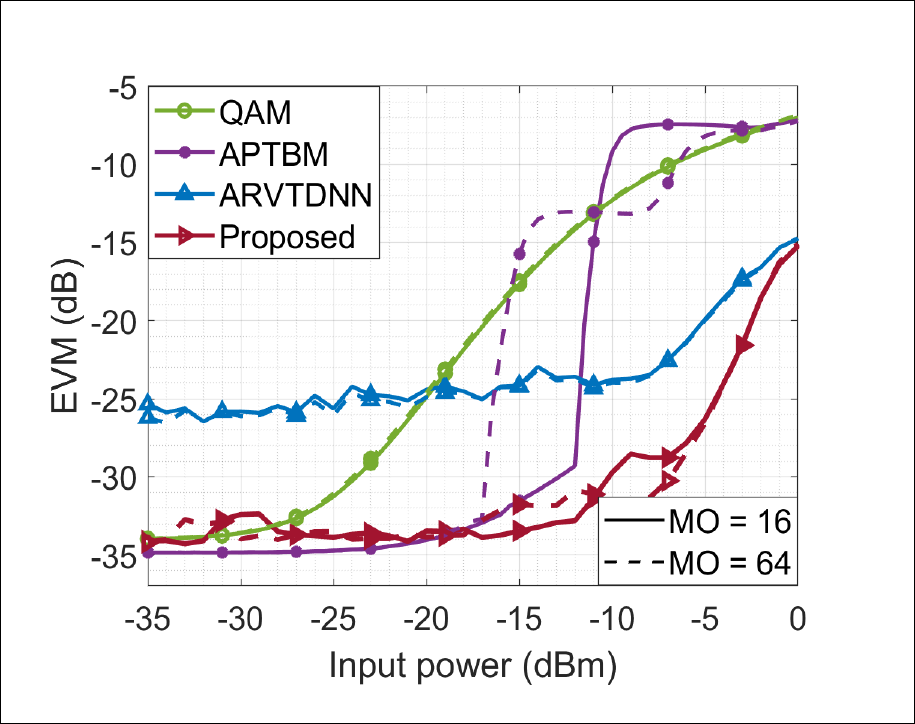}
    }
\subfigure[][]{
    \label{fig_sim_DPoD_diffSNR_BER_Saleh}
    \includegraphics[width=1.6in, trim=8mm 6mm 12mm 11mm, clip]{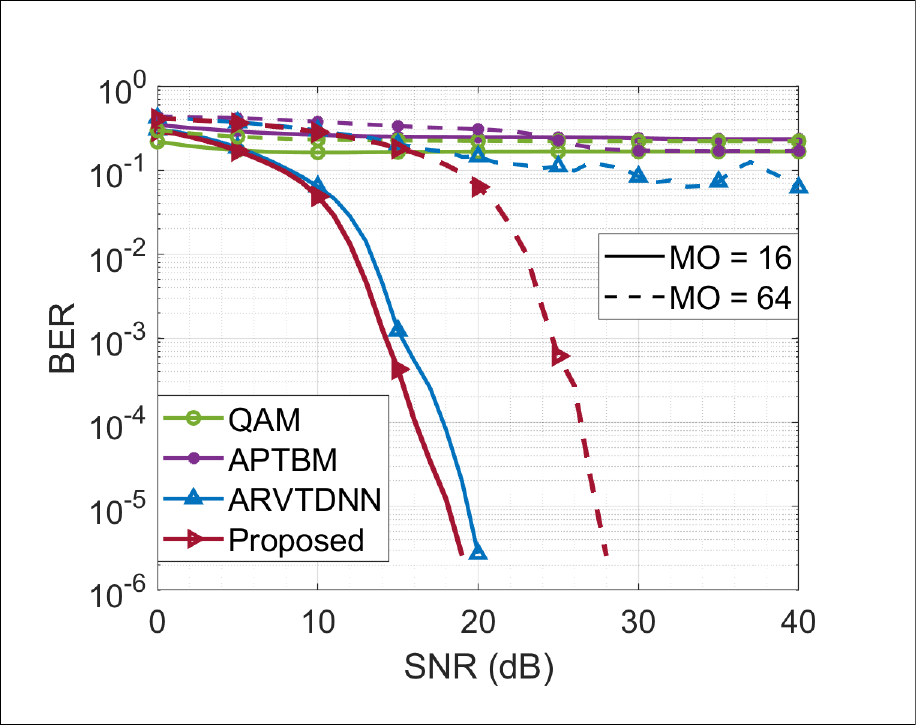}
    }
\subfigure[][]{
    \label{fig_sim_DPoD_diffInputPower_EVM_GMP}
    \includegraphics[width=1.6in, trim=8mm 6mm 12mm 12mm, clip]{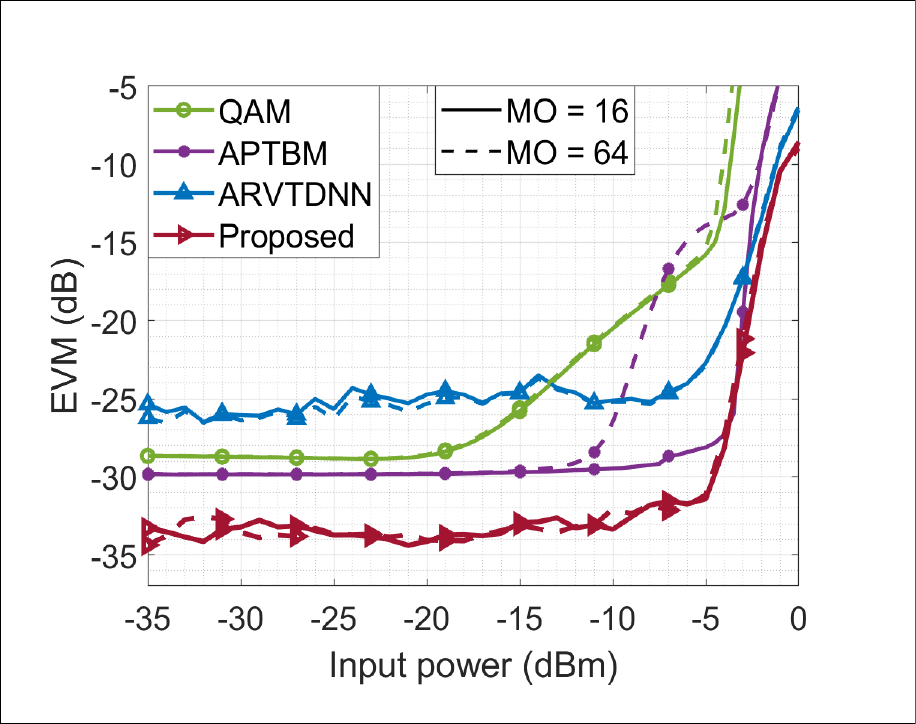}
    }
\subfigure[][]{
    \label{fig_sim_DPoD_diffSNR_BER_GMP}
    \includegraphics[width=1.6in, trim=8mm 6mm 12mm 11mm, clip]{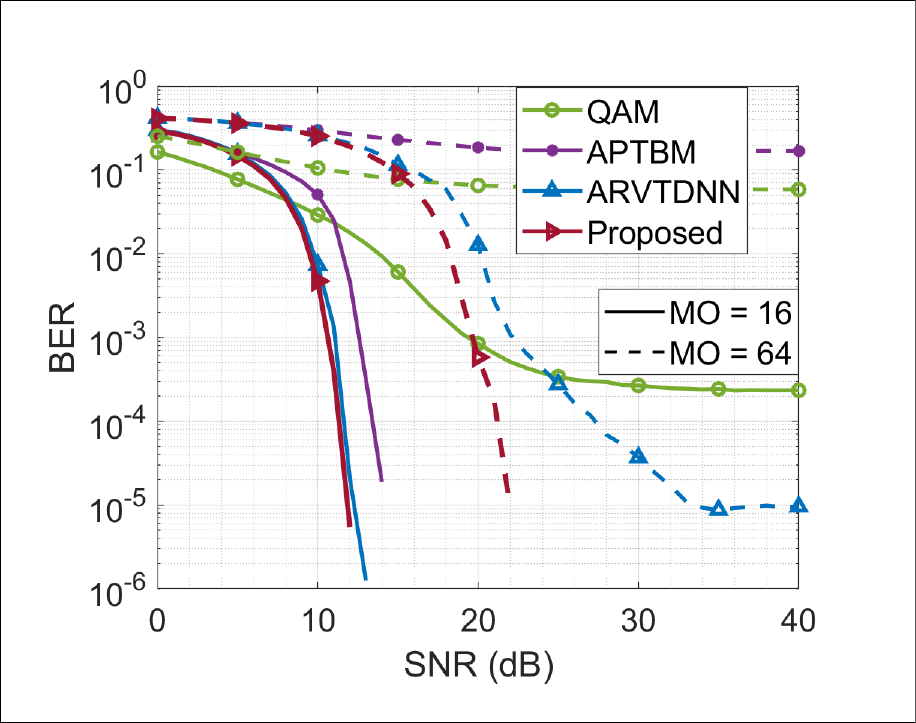}
    }
\caption{Simulation performance evaluation of the proposed DPoD scheme. (a)(b) Rapp model; (c)(d) Saleh model; (e)(f) GMP model.}
\label{fig_sim_diffInputPowers_diffSNR_EVM_BER}
\vspace{-6 pt}
\end{figure}

Fig. \ref{fig_sim_diffInputPowers_diffSNR_EVM_BER} compares the proposed DPoD scheme with quadrature
amplitude modulation (QAM), APTBM \cite{Fan2025_APTBM}, and ARVTDNN-based DPoD \cite{WangEtAl2019_ARVTDNN} under MO = $16$ and MO = $64$ configurations, with SNR = 30 dB, 100 offline/online training iterations, and $N_\mathrm{s}=1000$ ST symbols. As the PA input power increases, all schemes suffer from EVM degradation due to enhanced nonlinear distortion near saturation. APTBM exhibits performance degradation under severe AM-PM distortion, especially for the GMP model with memory effects, while ARVTDNN is limited by its higher training overhead. In contrast, the proposed scheme consistently achieves the lowest EVM across the Rapp, Saleh, and GMP models, benefiting from constraint-guided offline pretraining and efficient online adaptation. The BER results further confirm that the proposed scheme effectively suppresses the error floor observed in benchmark methods under strong nonlinear operation and high-order modulation.

\begin{figure}[tp]%
\centering
\subfigure[][]{
    \label{fig_DPoD_diffInputPowers_EVM_mea}
    \includegraphics[width=1.6in, trim=8mm 6mm 12mm 12mm, clip]{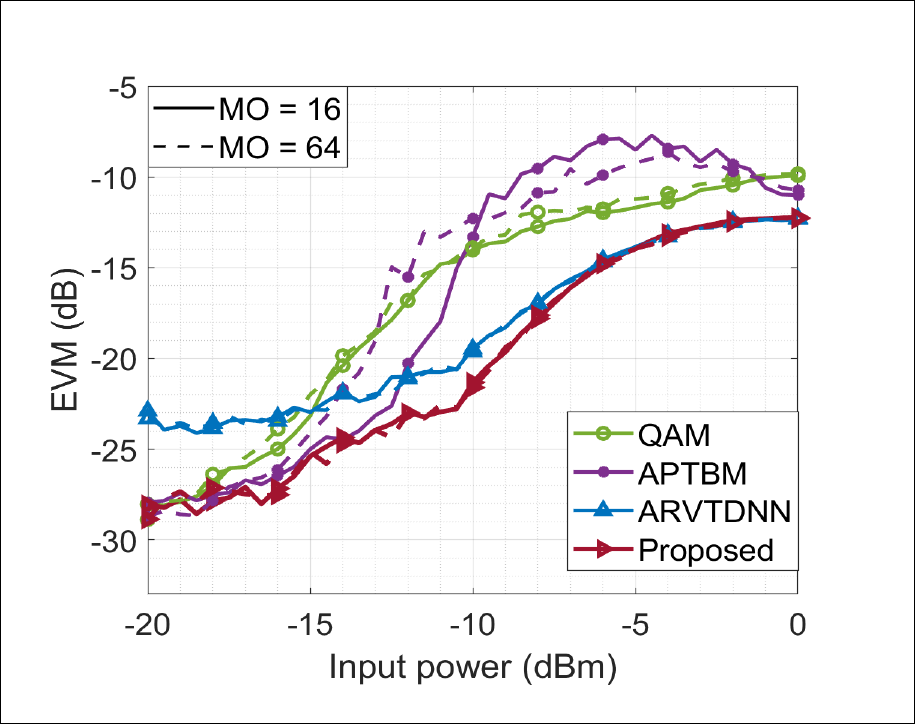}
    }
\subfigure[][]{
    \label{fig_DPoD_diffInputPowers_BER_mea}
    \includegraphics[width=1.6in, trim=8mm 6mm 12mm 12mm, clip]{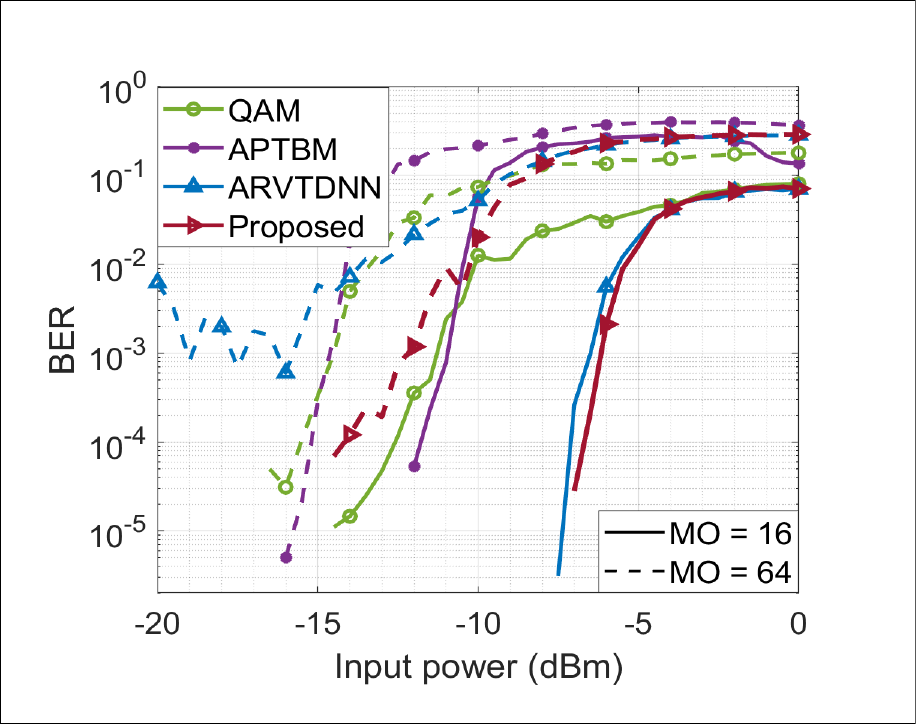}
    }
\caption{Measurement performance of the proposed DPoD scheme. (a) and (b) are the EVM and BER results, respectively.}
\label{fig_DPoD_diffInputPower_mea}
\vspace{-6 pt}
\end{figure}

\begin{figure}[tp]%
\centering
\subfigure[][]{
    \label{fig_diff_OnlineSTNumbers_EVM}
    \includegraphics[width=1.6in, trim=8mm 6mm 12mm 12mm, clip]{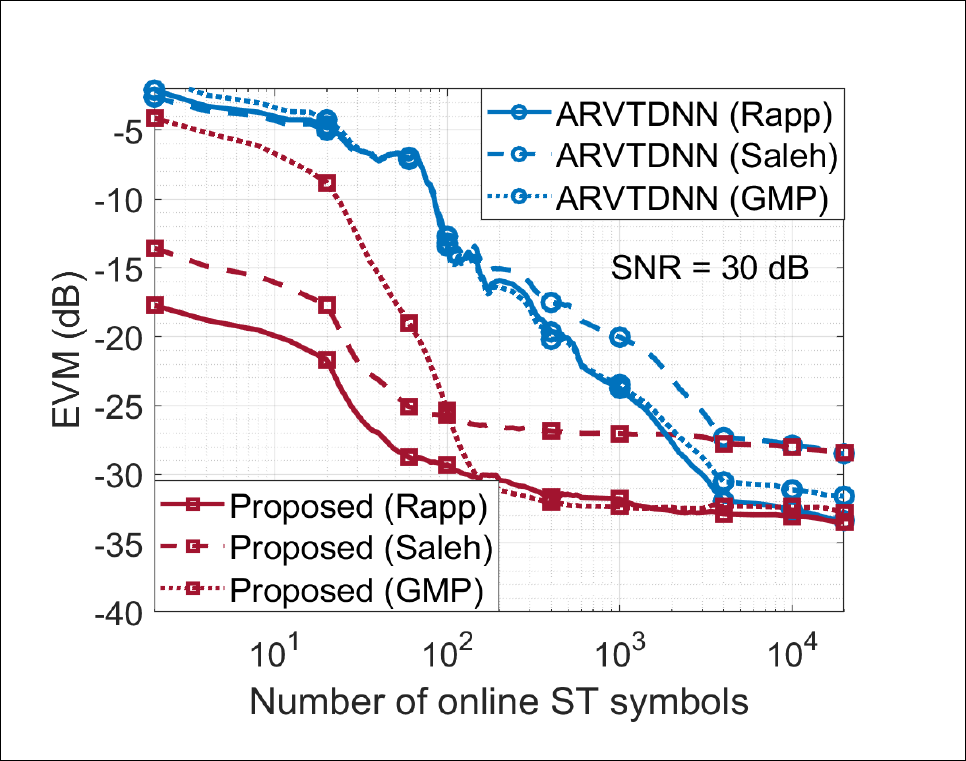}
    }
\subfigure[][]{
    \label{fig_diff_OnlineModelItes_EVM}
    \includegraphics[width=1.6in, trim=8mm 6mm 12mm 12mm, clip]{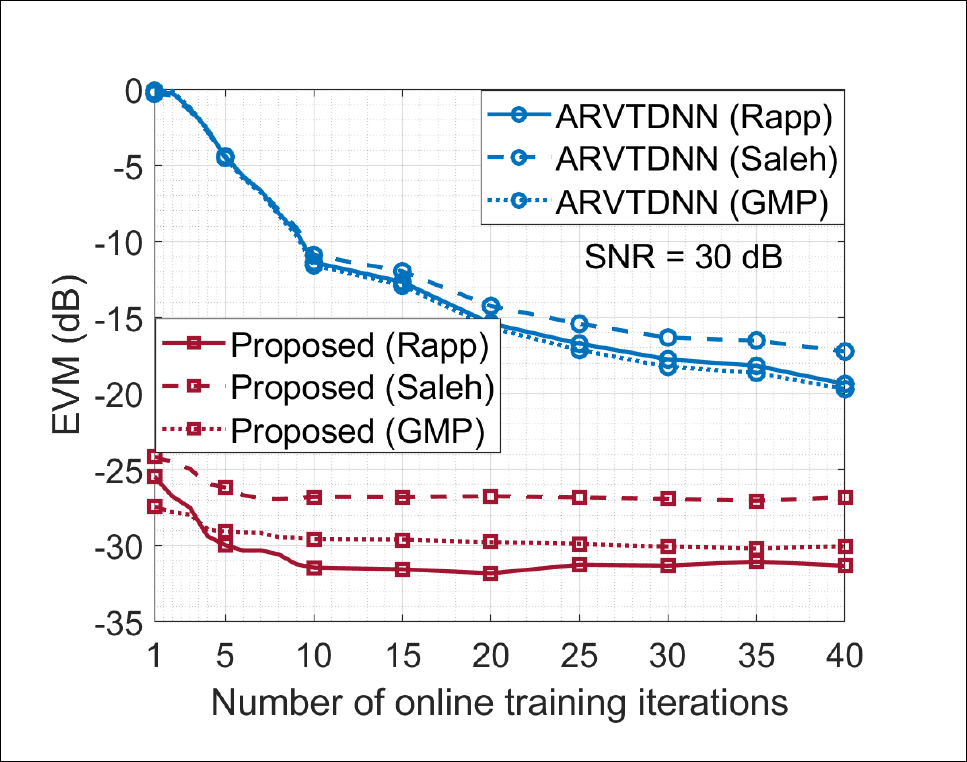}
    }
\subfigure[][]{
    \label{fig_diff_OnlineModelItes_MSE}
    \includegraphics[width=1.6in, trim=8mm 6mm 12mm 12mm, clip]{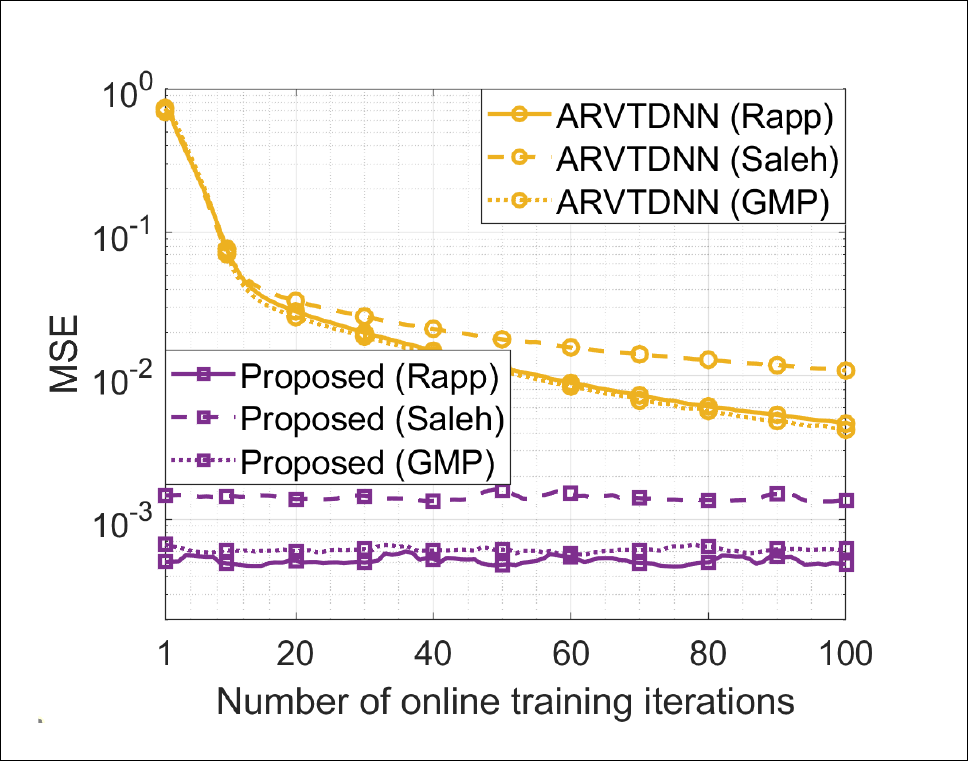}
    }
\subfigure[][]{
    \label{fig_diff_OnlineCNC_ite_EVM}
    \includegraphics[width=1.6in, trim=8mm 6mm 12mm 12mm, clip]{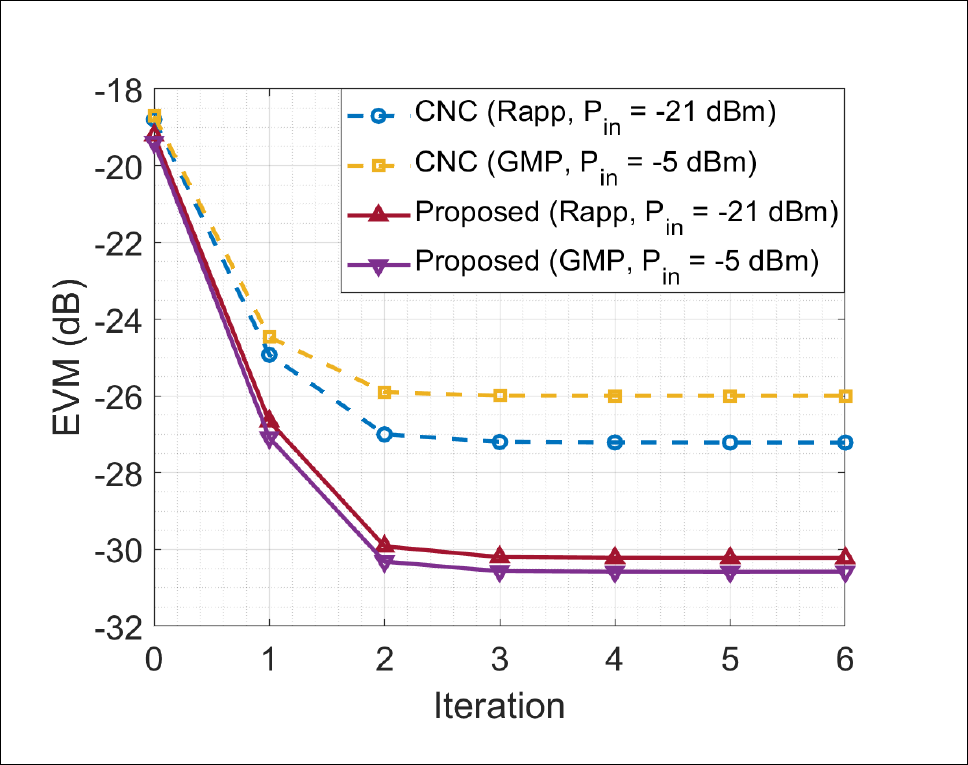}
    }
\caption{Online resource overhead analysis. (a) Compensation performance for different ST training sequence lengths; (b)(c) Compensation performance and convergence of online DPoD model training; (d) Online convergence of CNC algorithms.}
\label{fig_OnlineResourceOverhead_sim}
\vspace{-6 pt}
\end{figure}

The corresponding IBO and PAE improvements are further evaluated in Figs. \ref{fig_sim_DPoD_diffSNR_BER_Rapp}, \ref{fig_sim_DPoD_diffSNR_BER_Saleh}, and \ref{fig_sim_DPoD_diffSNR_BER_GMP}. At SNR = 30 dB, the operating input power is selected according to the minimum back-off required for reliable detection. For MOs of 16 and 64, the PA input powers are set to -7 and -11 dBm for the Rapp model, -2 and -4 dBm for the Saleh model, and -4 and -5 dBm for the GMP model, respectively. Compared with benchmark schemes, the proposed scheme reduces the required IBO by approximately $9\!-\!11$ dB, $3\!-\!8$ dB, and $10$ dB for the Rapp, Saleh, and GMP models, respectively. According to (\ref{Eq_IBO}) and (\ref{Eq_PAE}), these improvements correspond to PAE gains of $17.34\%\!-\!24.14\%$, $9.67\%\!-\!30.63\%$, and $23.79\%$, respectively.

The measured EVM and BER results in Fig. \ref{fig_DPoD_diffInputPower_mea} further validate the simulation trends. As the input power increases, APTBM suffers from rapid degradation due to intensified AM-PM distortion, whereas learning-based DPoD schemes extend the allowable operating range by compensating for PA nonlinearities. Under higher-order modulation, the proposed DPoD achieves superior robustness owing to constraint-aided offline knowledge and lightweight inverse modeling. At the same EVM level, it increases the allowable $P_{\mathrm{in}}$ by approximately 0.5 dB and 2 dB for MO = $16$ and MO = $64$, respectively, thereby enabling lower IBO operation and improved PA efficiency.

To further evaluate the online resource overhead, the proposed DPoD scheme is tested under Rapp, Saleh, and GMP PA models with input power levels of $-15$, $-10$, and $-5$ dBm, respectively. The SNR is set to 30 dB with MO = $64$. As shown in Fig. \ref{fig_diff_OnlineSTNumbers_EVM}, the proposed scheme achieves stable EVM performance with only approximately 100 online ST symbols for all PA models. At the same EVM level, it reduces the required online training data by about $20\times$ compared with ARVTDNN, owing to the offline pretrained initialization and efficient transfer learning-based adaptation.

Figs. \ref{fig_diff_OnlineModelItes_EVM} and \ref{fig_diff_OnlineModelItes_MSE} further evaluate the DPoD model convergence behavior under different online training iterations. Compared with conventional DPoD, the proposed scheme achieves faster convergence and improved steady-state compensation performance across different PA conditions. This gain results from the combination of weakly supervised pretraining and transfer learning-based adaptation, which enables efficient parameter updating and fast convergence with limited online training data. The proposed scheme achieves significant compensation improvement after only one iteration and maintains stable performance with approximately five iterations.

\subsection{ACLR-Compliant Compensation Results}
\label{SubSec_ACLRComplianceResults}

\begin{figure}[tp]%
\centering
\subfigure[][]{
    \label{fig_diff_ICAF_iterNums_CR_PAPR}
    \includegraphics[width=1.6in, trim=8mm 6mm 12mm 12mm, clip]{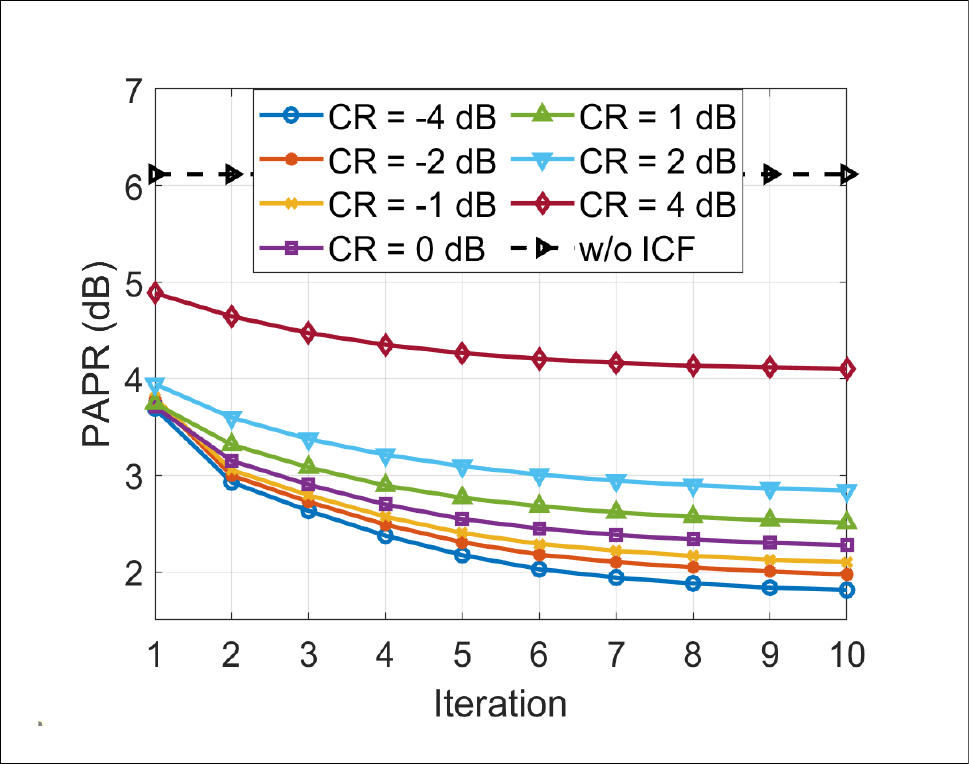}
    }
\subfigure[][]{
    \label{fig_diff_ICAF_iterNums_CR_EVM}
    \includegraphics[width=1.6in, trim=8mm 6mm 12mm 12mm, clip]{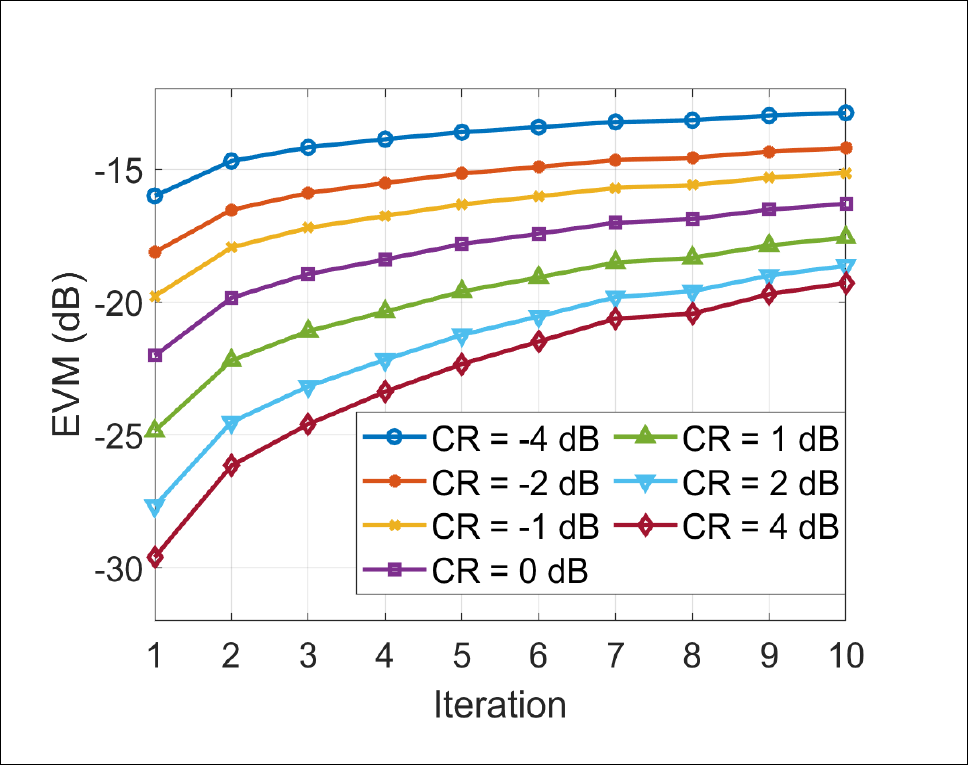}
    }
\subfigure[][]{
    \label{fig_diffInputPower_Tx_ACLR}
    \includegraphics[width=1.6in, trim=8mm 6mm 12mm 12mm, clip]{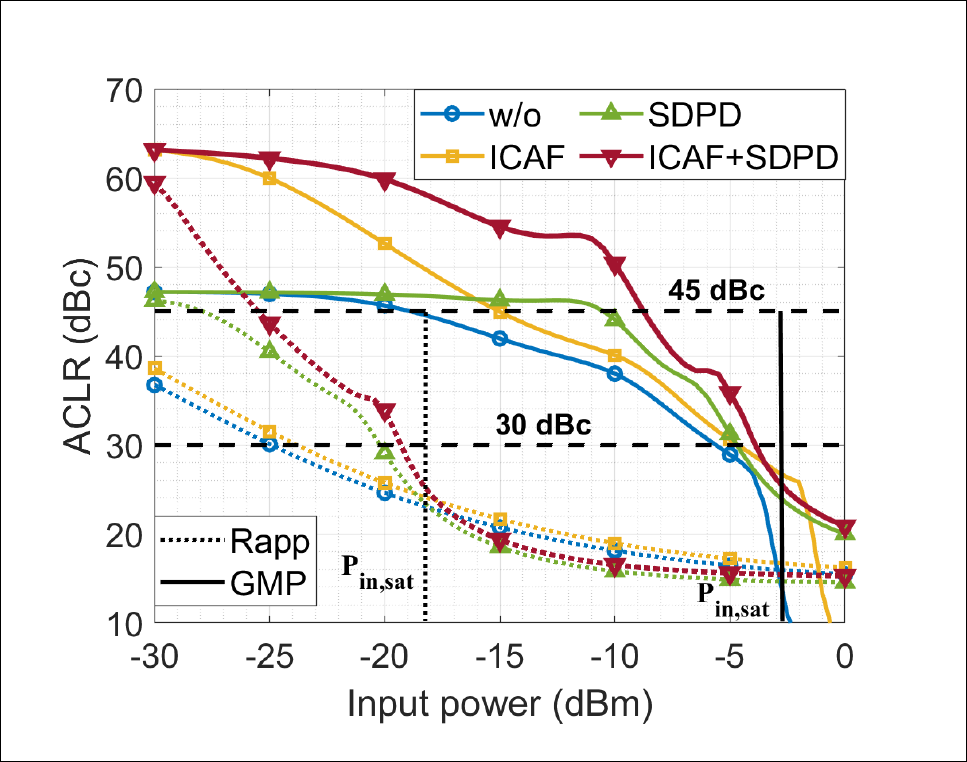}
    }
\subfigure[][]{
    \label{fig_diffInputPower_Tx_EVM}
    \includegraphics[width=1.6in, trim=8mm 6mm 12mm 12mm, clip]{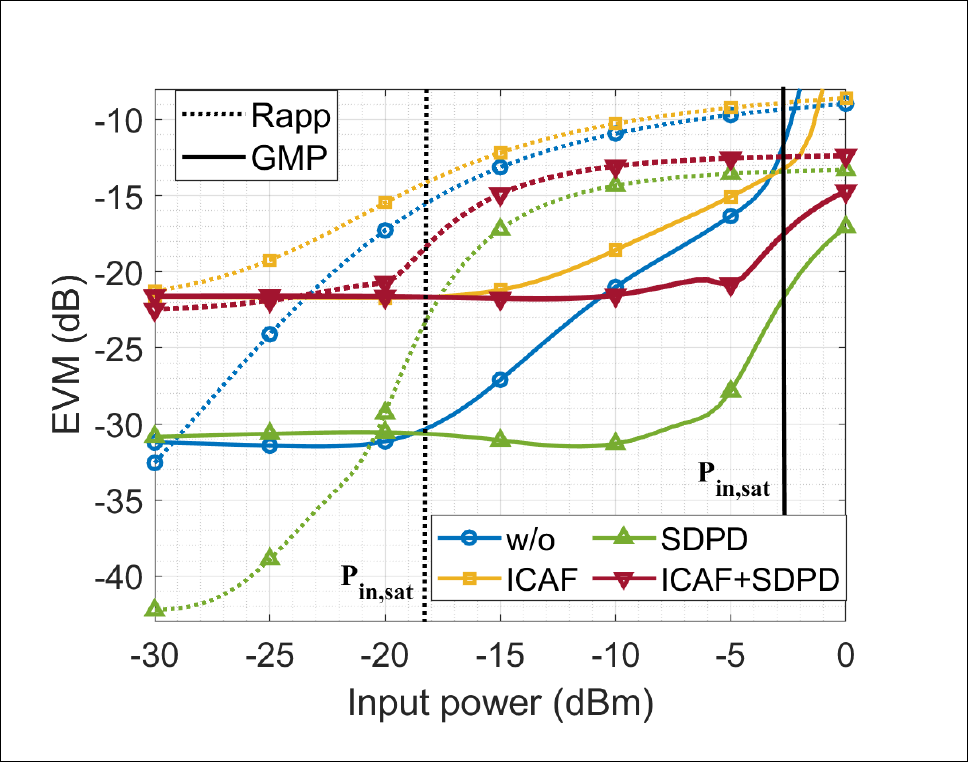}
    }
\subfigure[][]{
    \label{fig_diffInputPower_Tx_ACLR_mea}
    \includegraphics[width=1.6in, trim=8mm 6mm 12mm 12mm, clip]{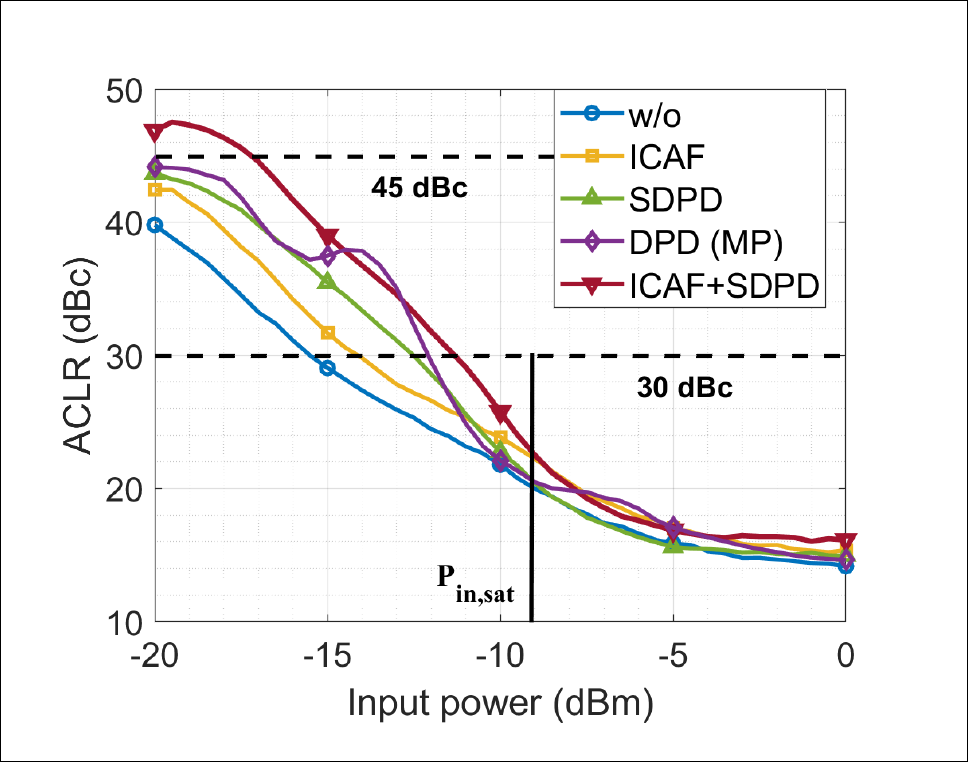}
    }
\subfigure[][]{
    \label{fig_diffInputPower_Tx_EVM_mea}
    \includegraphics[width=1.6in, trim=8mm 6mm 12mm 12mm, clip]{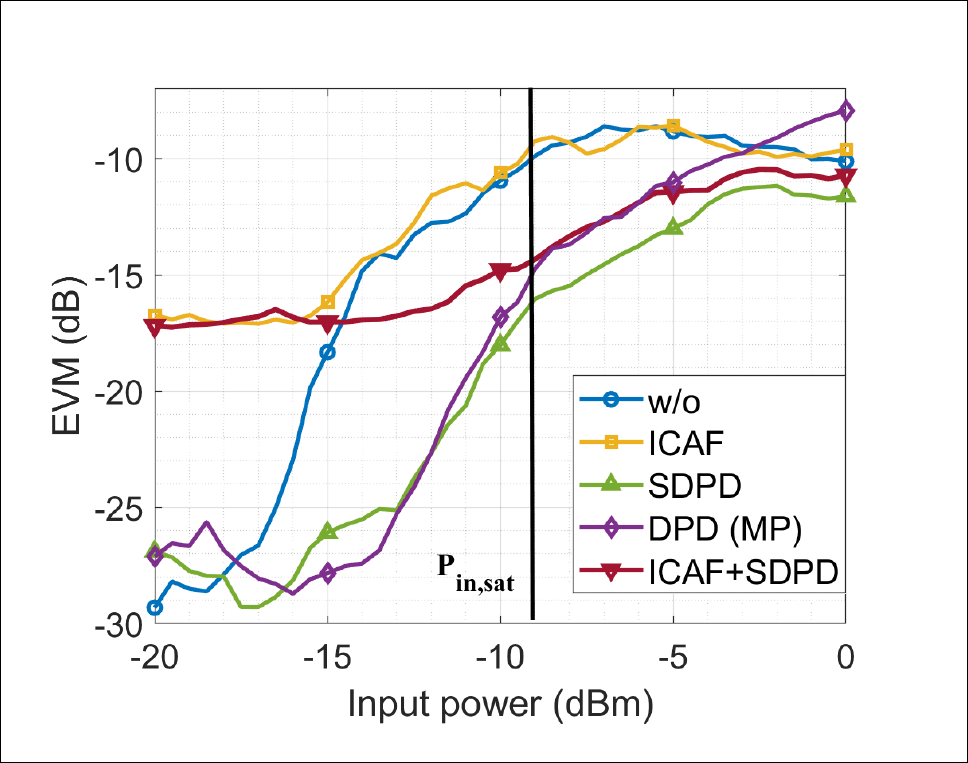}
    }
\caption{Transmitter processing results. (a) and (b) are the simulated PAPR and EVM of the ICAF; (c) and (d) are the simulated ACLR and EVM of the PA output after processing by the ICAF and SDPD; (e) and (f) are the measured ACLR and EVM for CR = 0 dB and one iteration of PA output.}
\label{fig_TXPerformanceAnalysis}
\vspace{-6 pt}
\end{figure}

In this section, the proposed transceiver-cooperative scheme is evaluated under ACLR constraints. Figs. \ref{fig_SDPD_AM_PM_all_sim} and \ref{fig_SDPD_test_PA_mea} first characterize the SDPD performance through simulations with the Rapp and GMP models and measurements using a practical PA. In both simulations and measurements, SDPD partially linearizes the AM-AM response and suppresses the amplitude-dependent AM-PM distortion over the usable operating range. In particular, the measured phase deviation is reduced from approximately $0.4$ rad to below $0.05$ rad, verifying effective static PA compensation without a wideband feedback path \cite{Fan2026_APFBM}. 

Additionally, Fig. \ref{fig_TXPerformanceAnalysis} evaluates the transmitter-side processing for MO = $64$. As shown in Figs. \ref{fig_diff_ICAF_iterNums_CR_PAPR} and \ref{fig_diff_ICAF_iterNums_CR_EVM}, stronger clipping and additional ICAF iterations reduce PAPR at the cost of increased in-band distortion. Without ICAF, the waveform PAPR is approximately $6.1$ dB. With $\mathrm{CR}=0$ dB and one iteration, it is reduced to about $3.8$ dB, while maintaining an EVM of approximately $-22$ dB. Increasing the iteration number to ten further reduces the PAPR to about $2.2$ dB but degrades the EVM to approximately $-17$ dB because of accumulated clipping distortion. Therefore, $\mathrm{CR}=0$ dB with one ICAF iteration is adopted as the PAPR and EVM trade-off for the subsequent experiments.

Figs. \ref{fig_diffInputPower_Tx_ACLR} and \ref{fig_diffInputPower_Tx_ACLR_mea} compare the ACLR performance of different transmitter schemes. ICAF limits large-envelope excursions, whereas SDPD suppresses PA-induced spectral regrowth. Under the $30$ dBc ACLR constraint, joint ICAF-SDPD processing increases the maximum permissible input power of the Rapp model from approximately $-25$ to $-19.5$ dBm, corresponding to an IBO reduction of about $5.5$ dB. Under the stricter $45$ dBc constraint, it supports $P_{\mathrm{in}}\approx-25.5$ dBm, approximately $2.5$ dB higher than SDPD alone, while the unprocessed and ICAF-only waveforms fail to satisfy the requirement over most of the considered range. For the GMP model, the corresponding IBO reductions under the $30$ and $45$ dBc constraints are approximately $2$ and $9.5$ dB, respectively.

\begin{figure}[tp]%
\centering
\subfigure[][]{
    \label{fig_Rx_diffInputPowers_EVM_sim}
    \includegraphics[width=1.6in, trim=8mm 6mm 12mm 12mm, clip]{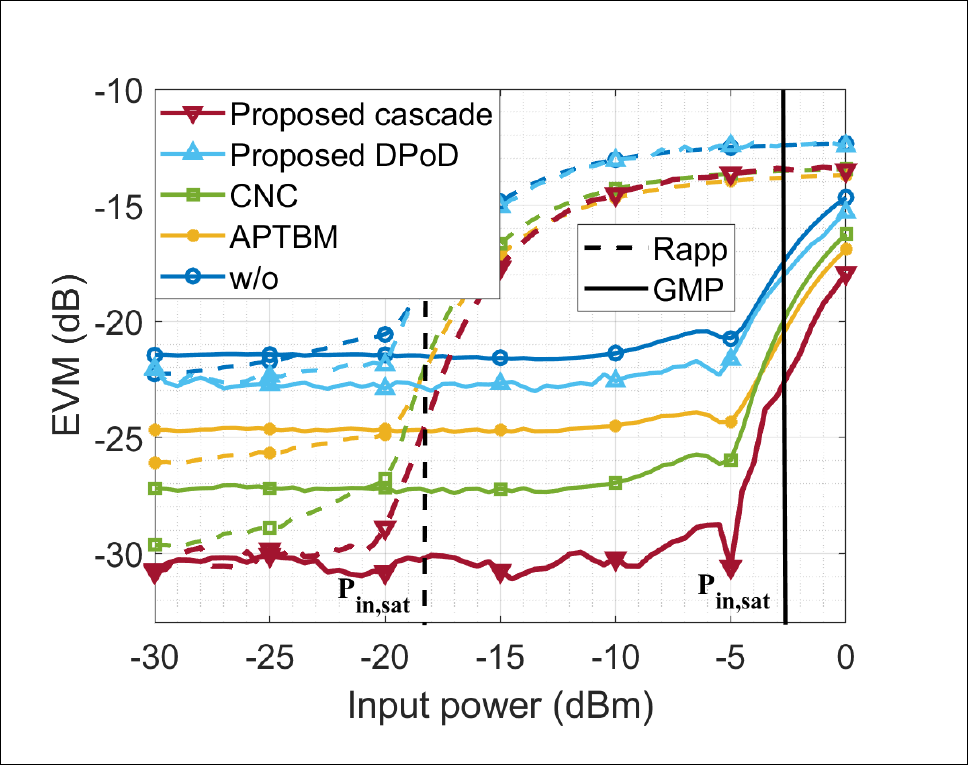}
    }
\subfigure[][]{
    \label{fig_Rx_diffInputPowers_BER_sim}
    \includegraphics[width=1.6in, trim=8mm 6mm 12mm 12mm, clip]{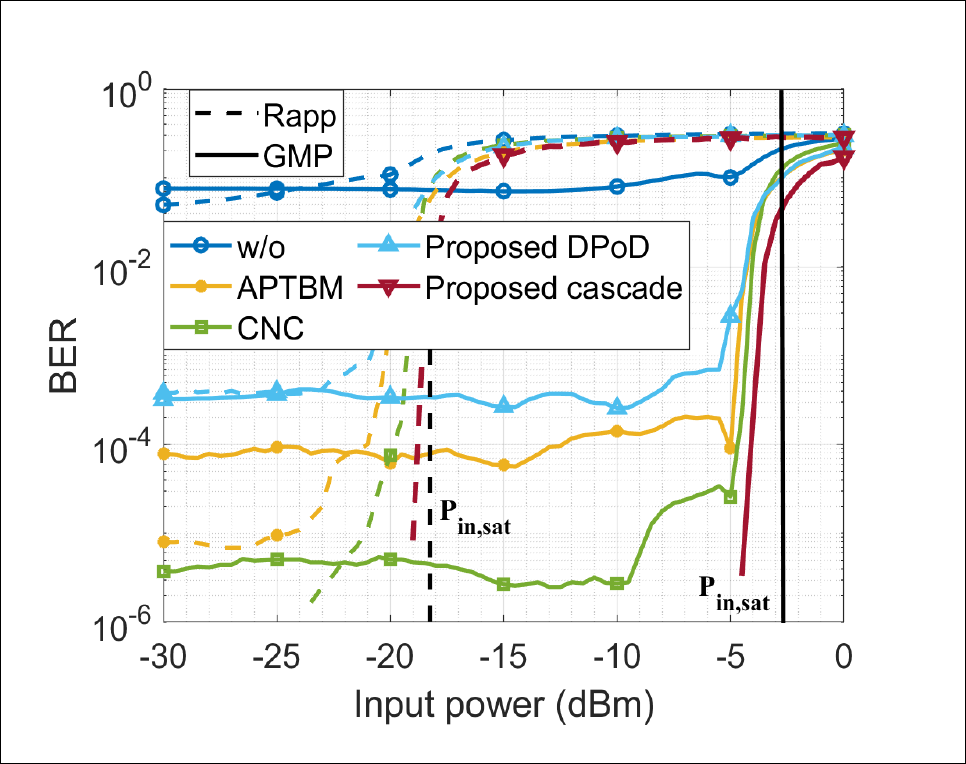}
    }
\subfigure[][]{
    \label{fig_Rx_diffInputPowers_EVM_mea}
    \includegraphics[width=1.6in, trim=8mm 6mm 12mm 12mm, clip]{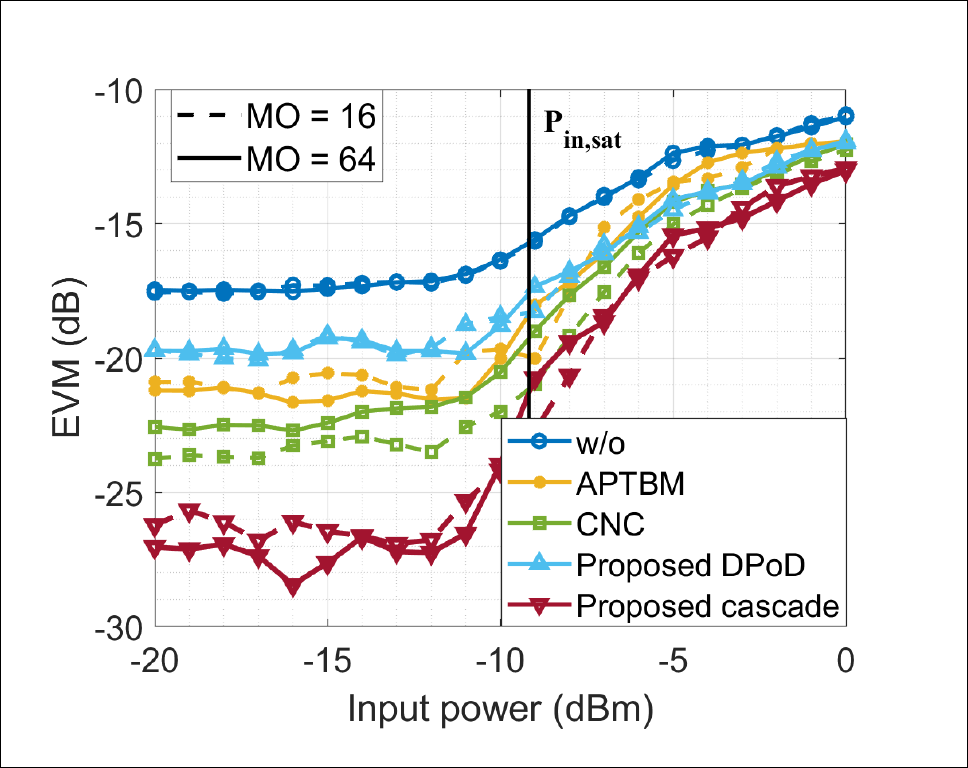}
    }
\subfigure[][]{
    \label{fig_Rx_diffInputPowers_BER_mea}
    \includegraphics[width=1.6in, trim=8mm 6mm 12mm 12mm, clip]{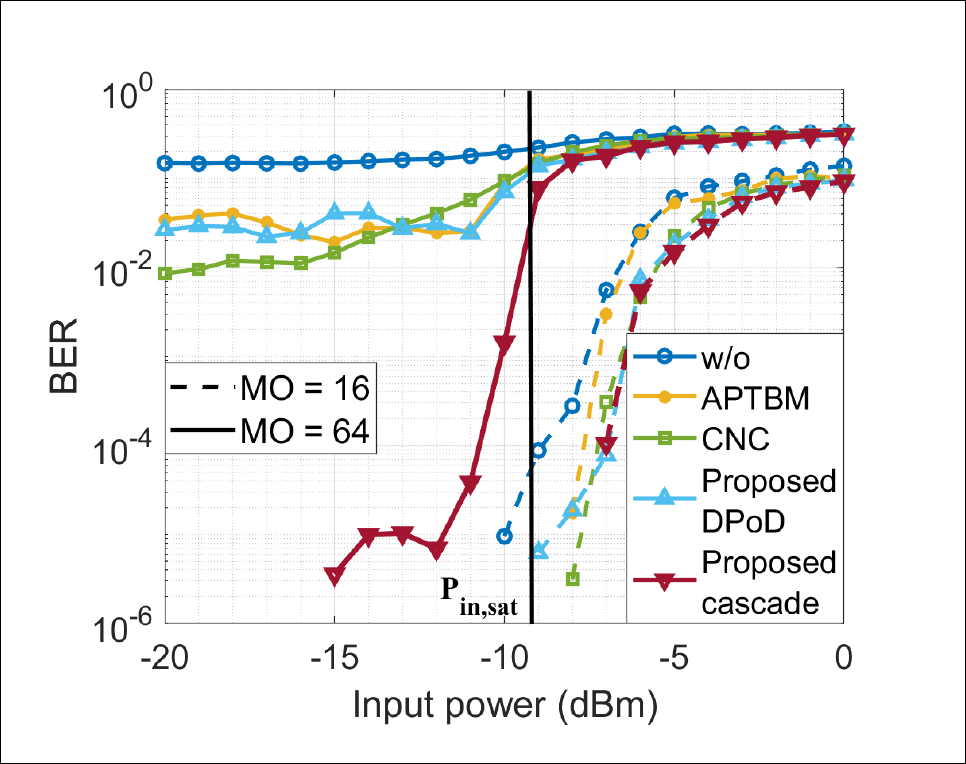}
    }
\subfigure[][]{
    \label{fig_Rx_diffCR_EVM_sim}
    \includegraphics[width=1.6in, trim=8mm 6mm 12mm 12mm, clip]{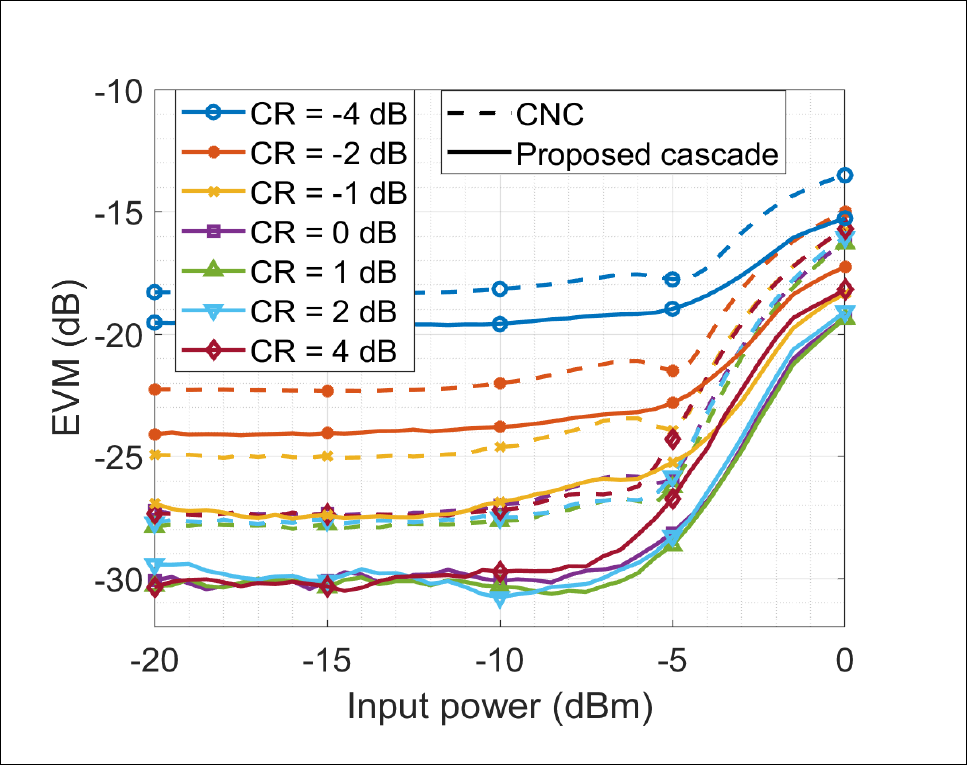}
    }
\subfigure[][]{
    \label{fig_Rx_diffIter_EVM_sim}
    \includegraphics[width=1.6in, trim=8mm 6mm 12mm 12mm, clip]{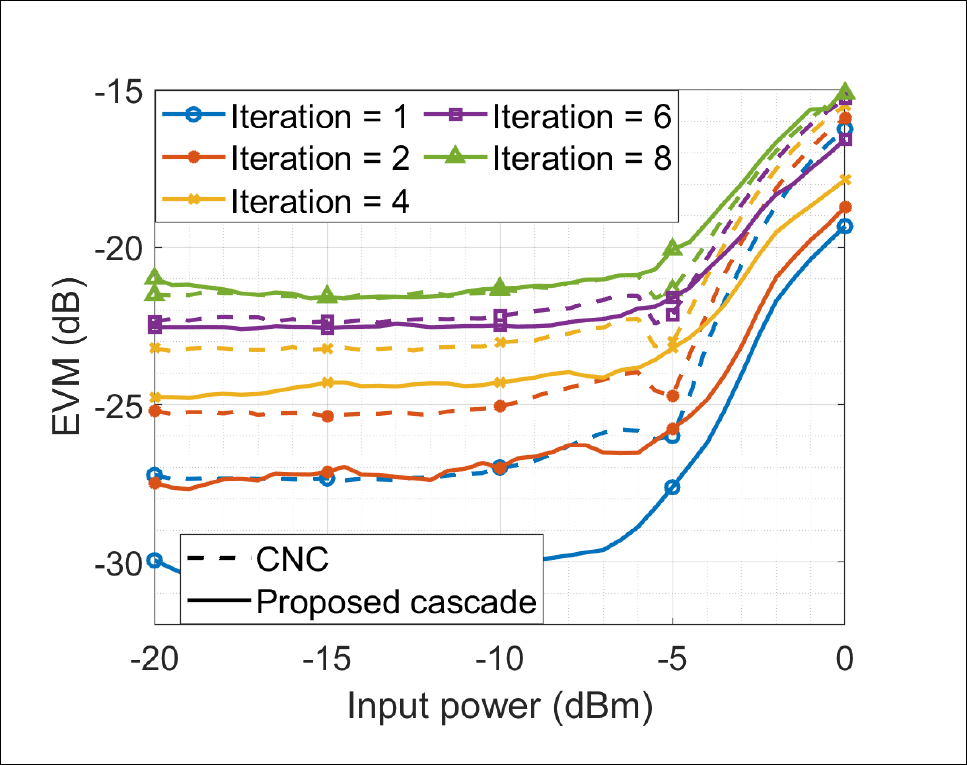}
    }
\caption{EVM and BER results for different compensation schemes in the receiver. (a) and (b) are simulation results for CR = 0 dB and one iteration; (c) and (d) are measured results for CR = 0 dB and one iteration; (e) and (f) are simulation results after compensation using different ICAF processing.}
\label{fig_RXPerformanceAnalysis}
\vspace{-6 pt}
\end{figure}

The measurements further compare ICAF-SDPD with MP-based DPD \cite{MorganEtAl2006_GMP}. Relative to MP-based DPD, the proposed processing reduces the required IBO by approximately $0.7$ dB under the $30$ dBc constraint and by more than $3$ dB under the $45$ dBc constraint; the reduction exceeds $4$ dB relative to the unprocessed waveform. Across the simulated and measured PAs, ICAF-SDPD enables an IBO of approximately $2$ dB, whereas the unprocessed baseline may require up to $7$ dB. The measured spectrum in Fig. \ref{fig_ACLR_PA_Output_mea} further confirms the reduced OOB leakage at $P_{\mathrm{in}}=-11$ dBm.

Figs. \ref{fig_diffInputPower_Tx_EVM} and \ref{fig_diffInputPower_Tx_EVM_mea} illustrate the associated in-band distortion. Although ICAF-SDPD substantially enlarges the ACLR-compliant operating range, its EVM is limited by the deterministic ICAF distortion that cannot be recovered by SDPD. In contrast, SDPD and MP-based DPD provide lower EVM under moderately nonlinear operation but allow for a smaller ACLR-compliant input power range. These results clarify the functional separation of the proposed architecture: transmitter processing ensures spectral compliance and low-IBO operation, while receiver processing recovers the PA and ICAF distortions.

Figs. \ref{fig_Rx_diffInputPowers_EVM_sim} and \ref{fig_Rx_diffInputPowers_BER_sim} compare the simulated receiver-side EVM and BER under $\mathrm{CR}=0$ dB and one transmitter ICAF iteration. DPoD alone suppresses the residual PA nonlinearity but retains the ICAF distortion, whereas conventional CNC reconstructs the clipping noise without fully compensating for the residual PA-induced distortion. The proposed DPoD-CNC cascade sequentially mitigates both distortions and therefore provides the lowest EVM and BER over the ACLR-compliant operating region. Under the $30$ dBc constraint, the maximum permissible input powers of the Rapp and GMP models are approximately $-19.5$ and $-4$ dBm, respectively. At these operating points, the proposed scheme improves EVM by at least $2-3$ dB over CNC and reduces BER below $10^{-6}$.

Using the same transceiver-side baseband processing as in the measurement setup, Figs. \ref{fig_Rx_diffInputPowers_EVM_mea} and \ref{fig_Rx_diffInputPowers_BER_mea} present the measured EVM and BER results for the different compensation schemes. Although hardware-state variations introduce moderate fluctuations, the measured trends closely agree with the simulations under joint ACLR and link-reliability constraints. Since SDPD compensates for the dominant PA distortion at the transmitter, nearly all benchmark schemes can operate close to saturation for MO = $16$. However, conventional CNC is noise-sensitive and cannot accurately reconstruct the ICAF distortion for high-order modulation, despite its competitive performance at MO = $16$. At a BER level on the order of $10^{-6}$, the proposed cascaded scheme reduces the required IBO by approximately $3$ dB for MO = $64$, thereby enabling operation at a substantially higher PA input power than the benchmark schemes.

Figs. \ref{fig_Rx_diffCR_EVM_sim} and \ref{fig_Rx_diffIter_EVM_sim} further evaluate robustness to different ICAF configurations using the GMP model. With one transmitter ICAF iteration, the proposed cascade consistently outperforms CNC over the considered CR range. With $\mathrm{CR}=0$ dB, increasing the ICAF iteration number progressively raises the clipping-distortion floor and degrades the receiver-side EVM. These results support a single transmitter ICAF iteration as a favorable compromise between peak suppression and signal fidelity. In addition, Fig. \ref{fig_diff_OnlineCNC_ite_EVM} characterizes receiver-side convergence. By first mitigating the residual PA nonlinearity, DPoD provides a cleaner waveform for clipping-noise reconstruction. Consequently, one iteration of the proposed cascade achieves approximately the EVM attained by conventional CNC after two iterations, and convergence is reached within about two receiver iterations. Further iterations provide only marginal improvement, confirming rapid convergence with limited receiver complexity.

\section{Conclusion}
\label{Sec_Con}

A transfer learning-enabled fully digital transceiver-cooperative framework has been proposed for ACLR-constrained nonlinear single-carrier wireless transmission. Transmitter-side ICAF-SDPD processing reduces PAPR and spectral regrowth, enabling efficient PA operation without wideband feedback, while the receiver exploits APTBM amplitude–phase constraints as structural priors to enable offline pretraining and online few-shot adaptation of a lightweight DPoD model. Furthermore, the cascaded DPoD and CNC scheme decouples residual PA nonlinearity from deterministic ICAF distortion, avoiding direct inversion of the coupled ICAF-PA response. Simulation and measurement results verify improved ACLR-compliant power efficiency, reliable high-order transmission, and fast receiver-side convergence under nonlinear and memory-dependent PA conditions. The proposed framework exploits APTBM-specific structural priors, motivating future exploration of more general constraints compatible with conventional QAM waveforms. Beyond single-carrier transmission, which is well suited for satellite and NTN applications, extending the proposed scheme to multicarrier systems is important to mitigate larger IBO requirements and further enhance PA efficiency.

\balance
\bibliographystyle{IEEEtran}
\bibliography{Refer}

\end{document}